\documentclass[11pt,a4paper]{article}
\usepackage{jinstpub}
\usepackage[sort&compress]{natbib}
\usepackage{lineno}

\usepackage{ragged2e}
\usepackage{amsmath,amsthm,amsfonts}
\usepackage{graphicx}
\usepackage{multirow}
\usepackage{subfigure}
\usepackage{afterpage}
\usepackage{float}
\usepackage{color}
\usepackage{placeins}
\definecolor{byz}{rgb}{0.74, 0.2, 0.64}

\newcommand{\LUT}{LUT} 
\newcommand{\BCR}{bunch-crossing reset}
\newcommand{\BC}{bunch-crossing}
\newcommand{\BCs}{bunch-crossings}
\newcommand{\DAQling}{{DAQling}} 
\newcommand{\Grafana}{{Grafana}}
\newcommand{\Redis}{{redis}}
\newcommand{\InfluxDB}{{InfluxDB}}

\begin{document}

\title{
{\Large The trigger and data acquisition system of the FASER experiment}
}

\collaboration{%
\includegraphics[height=17mm]{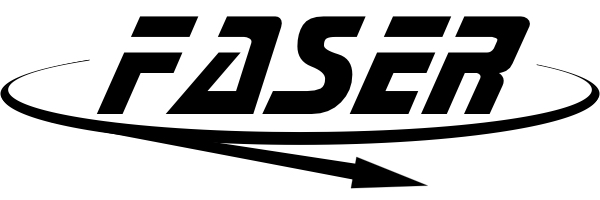}\\[6pt]
FASER collaboration }




\author[1]{Henso Abreu}

\author[2]{Elham Amin Mansour}

\author[2]{Claire Antel}

\author[3,21]{Akitaka Ariga}

\author[4]{Tomoko Ariga}

\author[5]{Florian Bernlochner}

\author[5]{Tobias Boeckh}

\author[6]{Jamie Boyd}

\author[6]{Lydia Brenner}

\author[2]{Franck Cadoux}

\author[7]{David~W.~Casper}

\author[8]{Charlotte Cavanagh}

\author[9]{Xin Chen}


\author[10]{Andrea Coccaro}

\author[2]{Stéphane Débieux}

\author[11]{Sergey Dmitrievsky}

\author[8]{Monica D’Onofrio}

\author[9]{Candan Dozen}


\author[2]{Yannick Favre}

\author[12]{Deion Fellers}

\author[7]{Jonathan~L.~Feng}

\author[2]{Didier Ferrere}

\author[6]{Enrico Gamberini}

\author[2]{Edward Karl Galantay}


\author[13]{Stephen Gibson}

\author[2]{Sergio Gonzalez-Sevilla}

\author[11]{Yuri Gornushkin}

\author[8]{Carl Gwilliam}

\author[14]{Shih-Chieh Hsu}

\author[9]{Zhen Hu}

\author[2]{Giuseppe Iacobucci}

\author[9]{Tomohiro Inada}

\author[6]{Sune Jakobsen}

\author[2]{Eliott Johnson}

\author[1]{Enrique Kajomovitz}

\author[15]{Felix Kling}

\author[6]{Umut Kose}

\author[6]{Susanne Kuehn}

\author[13]{Helena Lefebvre}

\author[16]{Lorne Levinson}

\author[14]{Ke Li}

\author[9]{Jinfeng Liu}

\author[2]{Chiara Magliocca}

\author[17]{Josh McFayden}

\author[2]{Matteo Milanesio}

\author[6]{Sam Meehan}

\author[6]{Dimitar Mladenov}

\author[2]{Théo Moretti}

\author[2]{Magdalena Munker}

\author[18]{Mitsuhiro Nakamura}

\author[18]{Toshiyuki Nakano}

\author[6]{Marzio Nessi}

\author[19]{Friedemann Neuhaus}

\author[13]{Laurie Nevay}

\author[4]{Hidetoshi Otono}


\author[2]{Carlo Pandini}

\author[9]{Hao Pang}

\author[2]{Lorenzo Paolozzi}

\author[6]{Brian Petersen}

\author[6]{Francesco Pietropaolo}

\author[5]{Markus Prim}

\author[6]{Michaela Queitsch-Maitland}

\author[6]{Filippo Resnati}

\author[2]{Chiara Rizzi}

\author[18]{Hiroki Rokujo}

\author[19]{Elisa Ruiz-Choliz}

\author[6]{Jakob Salfeld-Nebgen}

\author[18]{Osamu Sato}

\author[3,22]{Paola Scampoli}

\author[19]{Kristof Schmieden}

\author[19]{Matthias Schott}

\author[2]{Anna Sfyrla}

\author[7]{Savannah Shively}

\author[6]{Roland Sipos}


\author[14]{John Spencer}

\author[20]{Yosuke Takubo}

\author[2]{Noshin Tarannum}

\author[2]{Ondrej Theiner}

\author[12]{Eric Torrence}


\author[6]{Serhan Tufanli}

\author[11]{Svetlana Vasina}

\author[6]{Benedikt Vormwald}

\author[9]{Di Wang}



\affiliation[1]{Department of Physics and Astronomy, Technion---Israel Institute of Technology, Haifa 32000, Israel}

\affiliation[2]{D\'epartement de Physique Nucl\'eaire et Corpusculaire, 
University of Geneva, CH-1211 Geneva 4, Switzerland}


\affiliation[3]{Albert Einstein Center for Fundamental Physics, Laboratory for High Energy Physics, University of Bern, Sidlerstrasse 5, CH-3012 Bern, Switzerland}

\affiliation[4]{Kyushu University, Nishi-ku, 819-0395 Fukuoka, Japan}

\affiliation[5]{Universit\"at Bonn, Regina-Pacis-Weg 3, D-53113 Bonn, Germany}

\affiliation[6]{CERN, CH-1211 Geneva 23, Switzerland}

\affiliation[7]{Department of Physics and Astronomy, 
University of California, Irvine, CA 92697-4575, USA}

\affiliation[8]{University of Liverpool, Liverpool L69 3BX, United Kingdom}

\affiliation[9]{Department of Physics, Tsinghua University, Beijing, China}

\affiliation[10]{INFN Sezione di Genova, Via Dodecaneso, 33--16146, Genova, Italy}

\affiliation[11]{Joint Institute for Nuclear Research, Dubna, Russia}



\affiliation[12]{University of Oregon, Eugene, OR 97403, USA}


\affiliation[13]{Royal Holloway, University of London, Egham, TW20 0EX, UK}

\affiliation[14]{Department of Physics, University of Washington, PO Box 351560, Seattle, WA 98195-1560, USA}

\affiliation[15]{Deutsches Elektronen-Synchrotron DESY, Notkestrasse 85, 22607 Hamburg, Germany}

\affiliation[16]{Department of Particle Physics and Astrophysics, Weizmann Institute of Science, Rehovot 76100, Israel}

\affiliation[17]{Department of Physics \& Astronomy, University of Sussex, Sussex House, Falmer, Brighton, BN1 9RH, United Kingdom}


\affiliation[18]{Nagoya University, Furo-cho, Chikusa-ku, Nagoya 464-8602, Japan}

\affiliation[19]{Institut f\"ur Physik, Universität Mainz, Mainz, Germany}

\affiliation[20]{Institute of Particle and Nuclear Study, 
KEK, Oho 1-1, Tsukuba, Ibaraki 305-0801, Japan}


\affiliation[21]{Department of Physics, Chiba University, 1-33 Yayoi-cho Inage-ku, Chiba, 263-8522, Japan}

\affiliation[22]{Dipartimento di Fisica ``Ettore Pancini'', Universit\`a di Napoli Federico II, Complesso Universitario di Monte S. Angelo, I-80126 Napoli, Italy}

\abstract{
The FASER experiment is a new small and inexpensive experiment that is placed 480 meters downstream of the ATLAS experiment at the CERN LHC. FASER is designed to capture decays of new long-lived particles, produced outside of the ATLAS detector acceptance. These rare particles can decay in the FASER detector together with about 500--1000~Hz of other particles originating from the ATLAS interaction point. A  very  high  efficiency trigger and data acquisition system is required to ensure that the physics events of interest  will be recorded. This paper describes the trigger and  data acquisition system of the FASER experiment and presents  performance  results of the  system  acquired during initial commissioning.
}

\maketitle

\clearpage




\section{The FASER experiment at the CERN LHC }\label{sec:intro}

\begin{figure}[H]
    \centering
    \includegraphics[width=0.75\textwidth]{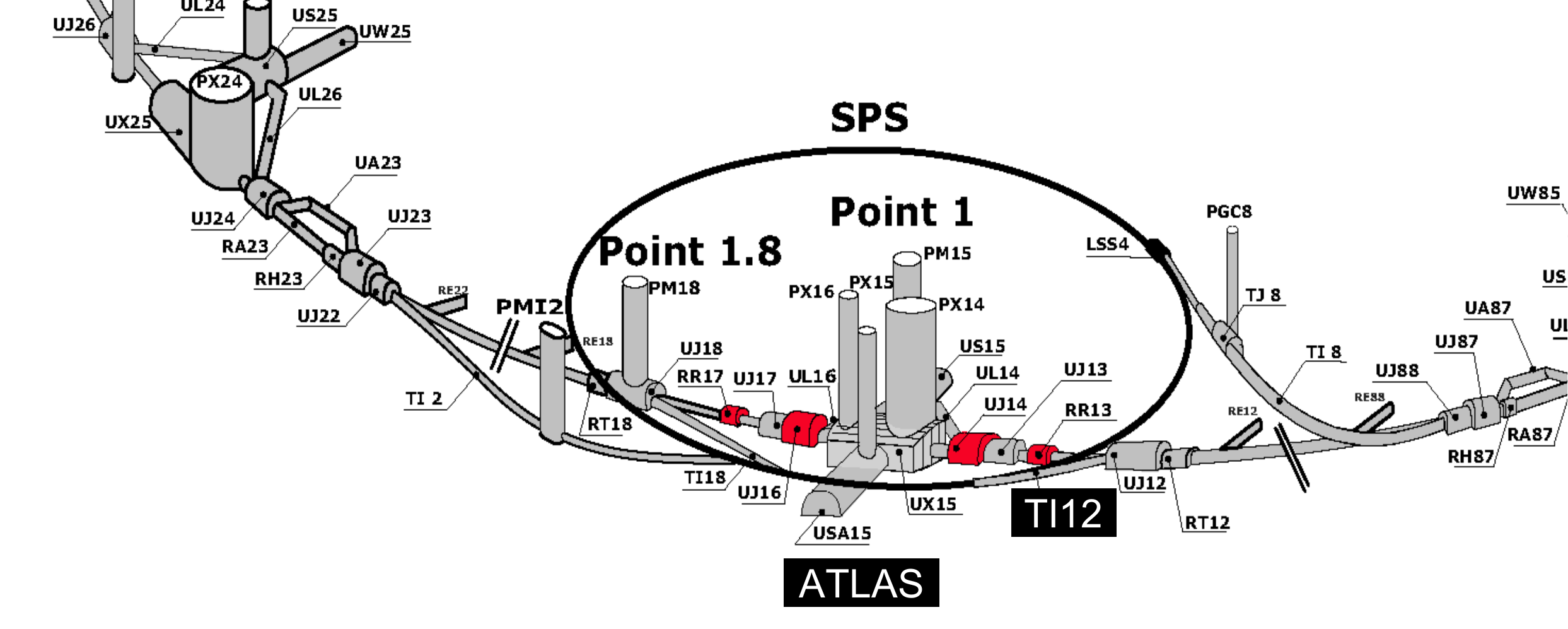}
    \caption{The FASER location: TI12 tunnel, 480~m downstream of the ATLAS interaction point. The detector is located along the beam collision axis line of sight.}
    \label{fig:FASER_location}
\end{figure}

The FASER experiment~\cite{FASERLoI, FASERTP} is a new small experiment located along the beam collision axis line of sight (LOS), about 480~m downstream of ATLAS, in the TI12 tunnel in the LHC complex at CERN, as shown in figure~\ref{fig:FASER_location}. The FASER detector design is motivated by the signature of a new particle, such as a dark photon, decaying to two  high-energy oppositely charged particles, such as an electron-positron pair~\cite{Feng:2017uoz,FASER_LLP}. For a dark photon of 100~MeV mass and 1~TeV energy,  the two particles to which it will decay are expected to be extremely collimated; this leads to requirements for a strong magnetic field and excellent tracking resolution, such that the electron and positron can be identified, as well as good energy measurements for electromagnetic deposits. The detector design has as additional drivers the cost and tight timeline between experiment approval (March 2019) and desired installation (before the Run 3 data taking would start).  This results in a detector that recycles active components from other experiments and minimizes services to simplify installation and operation. About 500--1000~Hz of energetic particles from the interaction point (IP) will be recorded by the trigger and data acquisition (TDAQ) system of the experiment during proton--proton ($pp$) collisions in ATLAS. Despite its simple design, the TDAQ system necessitates robustness and stability to guarantee high efficiency in recording signal events, which are expected to be rare.  

\begin{figure}[h]
    \centering
    \includegraphics[trim=0cm 0.8cm 1cm 4.5cm, clip=true, width=\textwidth]{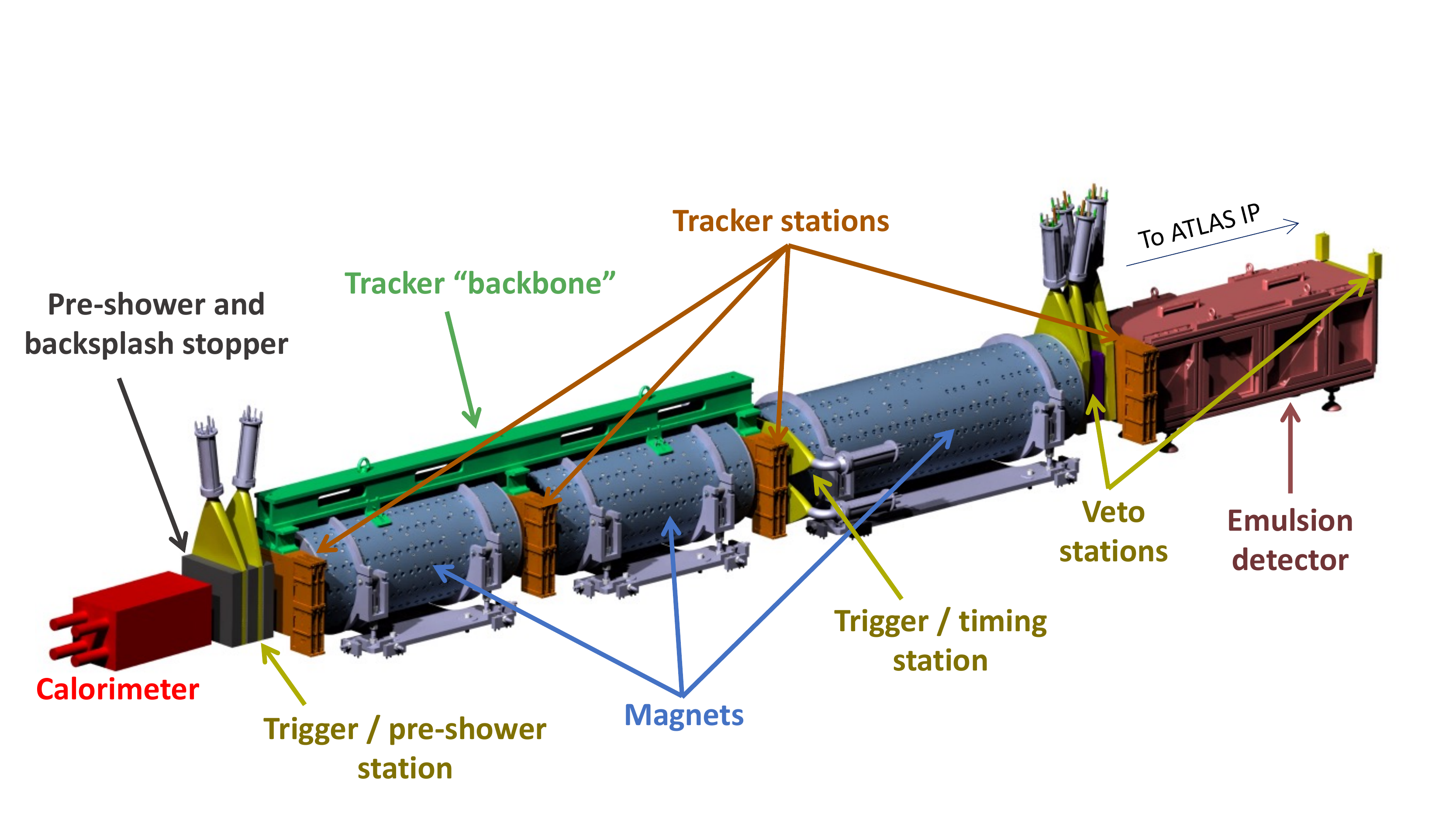}
    \caption{A CAD drawing of the FASER  detector.}
    \label{fig:FASER_labels}
\end{figure}

The FASER detector, all components of which are shown in figure~\ref{fig:FASER_labels}, is a 5.5~m-long spectrometer, with a 20~cm aperture and a decay volume length of 1.5~m.  The detector consists of tracking stations, scintillator stations and a calorimeter. It is complemented by additional detector components that constitute the FASER$\nu$ part of the experiment~\cite{FASERnuLoI,FASERnuTP}, which aims at precisely measuring cross-sections of neutrinos originating from the ATLAS IP. FASER$\nu$ is composed of a 1.1~tn emulsion film detector with tungsten plates and is complemented by an additional tracker station, referred to as the interface tracker station, IFT, and scintillators. This part of the detector was proposed and approved later than the rest.

The FASER detector volume is dominated by three dipole permanent magnets of a fixed 0.55~T field with Samarium Cobalt (Sm$_2$Co$_{17}$) magnetic material. They are based on a Halbach array design~\cite{HALBACH19801}, which maximizes the field without the need for complicated support infrastructure. The first of the three magnets has a length of 1.5~m and surrounds the  experiment's decay volume. The other two magnets are each 1~m long. 

The tracker is a key component of  the FASER detector. It is composed of four tracking stations of three silicon layers each. Each silicon layer is composed of 8 double-sided silicon strip modules, resulting in 96 modules in total. ATLAS SCT~\cite{SCT} spare modules are used, carefully selected based on a series of quality assurance tests. The modules are assembled in aluminum frames and read out using general purpose readout boards. A mechanical structure connecting the three stations and the two magnets that make up the FASER spectrometer, guarantees the relative alignment between these tracking stations, while the IFT is mechanically attached to the detector frame with a separate support structure. For smooth operations, the front-end electronics are kept at a stable temperature using a water chiller placed in TI12 close to the detector. 

Four scintillator stations provide trigger signals. The first of the stations is located in front of the FASER$\nu$ emulsion detector. The second is located in front of the first magnet and is composed of four layers with an absorber block of lead between the first two and the last two. Each layer has a detection efficiency above 99.9\% for minimum ionizing particles, resulting in a very high veto efficiency for any muon-induced background. These first two stations are referred to as the `veto stations'. The third station, or `timing station', is located on the other end of the first magnet and is used to detect the presence of charged particles in the detector. It also provides a  measurement of the arrival time of any charged signal with respect to the $pp$ interaction at the ATLAS IP to a precision of less than 1~ns. The fourth scintillator station, or `preshower station',  is located in front  of the  calorimeter. Two scintillators are interleaved by graphite blocks to minimize back-scatter of low energy particles from the calorimeter and 3\,mm tungsten layers to create a simple pre-shower detector that can help identify a physics signal of energetic photons. The veto and pre-shower stations use 2 cm thick, 30~cm$\times$30~cm plastic scintillators, while the timing station employs two 1-cm thick, 20~cm$\times$40~cm blocks  arranged to form a 40~cm$\times$40~cm single layer. All scintillators are connected to photomultiplier tubes (PMTs) and are read out via a commercial CAEN digitizer board.

Triggering capability is also provided by the calorimeter, which is otherwise used to identify  electrons and photons and to measure their energies. Four spare LHCb outer  electromagnetic calorimeter~\cite{LHCbECAL} modules are used. They provide energy resolution of about 1\% for 100\,GeV to TeV energy deposits. They have a total depth of 25 radiation lengths, but with no longitudinal shower information. They are connected to PMTs, which are read out via the  same digitizer board used for the scintillators. 

A LED-based system is used for calibrating and monitoring the stability of the response of the calorimeter modules and scintillators. It provides short light pulses of the order of 10~ns with variable amplitude sent via optical fibers to the calorimeter modules or scintillators. The calibration is performed by producing signals and measuring the gain as a function of the PMT bias voltage.

The tracker readout boards and the CAEN digitizer board are described in section~\ref{sec:TDAQ_global}, together with other general aspects of the FASER's TDAQ system. The boards are described in more detail and together with the rest of the TDAQ hardware components in section~\ref{sec:trigger_hw}. The power supplies, chiller, temperatures (both of detector components and the environment) and humidity are controlled and/or monitored using a dedicated detector control system (DCS). The TDAQ software components (including for the monitoring and DCS) are described in the subsequent sections~\ref{sec:daqsoftware} and~\ref{sec:controls_monitoring}.

The FASER detector is being tested and commissioned in three phases: (a) individual detector pieces and electronics cards were tested in dedicated labs at the CERN Meyrin site or the University of Geneva; (b) large detector pieces assembled  in groups, including a complete TDAQ chain, were tested in a dedicated surface area at the CERN Prevessin site; (c) the full detector, except for the components corresponding to FASER$\nu$, was assembled in the TI12 tunnel by March 2021 and full-system tests have been taking place since. All testing steps have used either generator pulses or cosmic muons. The FASER$\nu$ installation and commissioning are envisaged for autumn 2021. The aim is to have the full detector ready in early 2022 for commissioning with beams and subsequent data taking. 
This paper includes commissioning results of the TDAQ system in section~\ref{sec:standalone_commissioning}, focusing the discussion to TDAQ aspects only. Sections~\ref{sec:digitizer_com}--\ref{sec:TRB_com} describe the individual component testing and commissioning, section~\ref{sec:DAQ_com} describes the DAQ software and server performance tests and section~\ref{sec:combined_com} presents full-system commissioning results. 
The FASER TDAQ system described in this paper is sized to that required for the full detector, including the tracker and scintillator stations corresponding to the FASER$\nu$ part of the experiment.

\section{TDAQ system overview}\label{sec:TDAQ_global}

\begin{figure}[h]
    \centering
    \includegraphics[trim=1cm 0cm 9cm 0cm, clip=true, width=\textwidth]{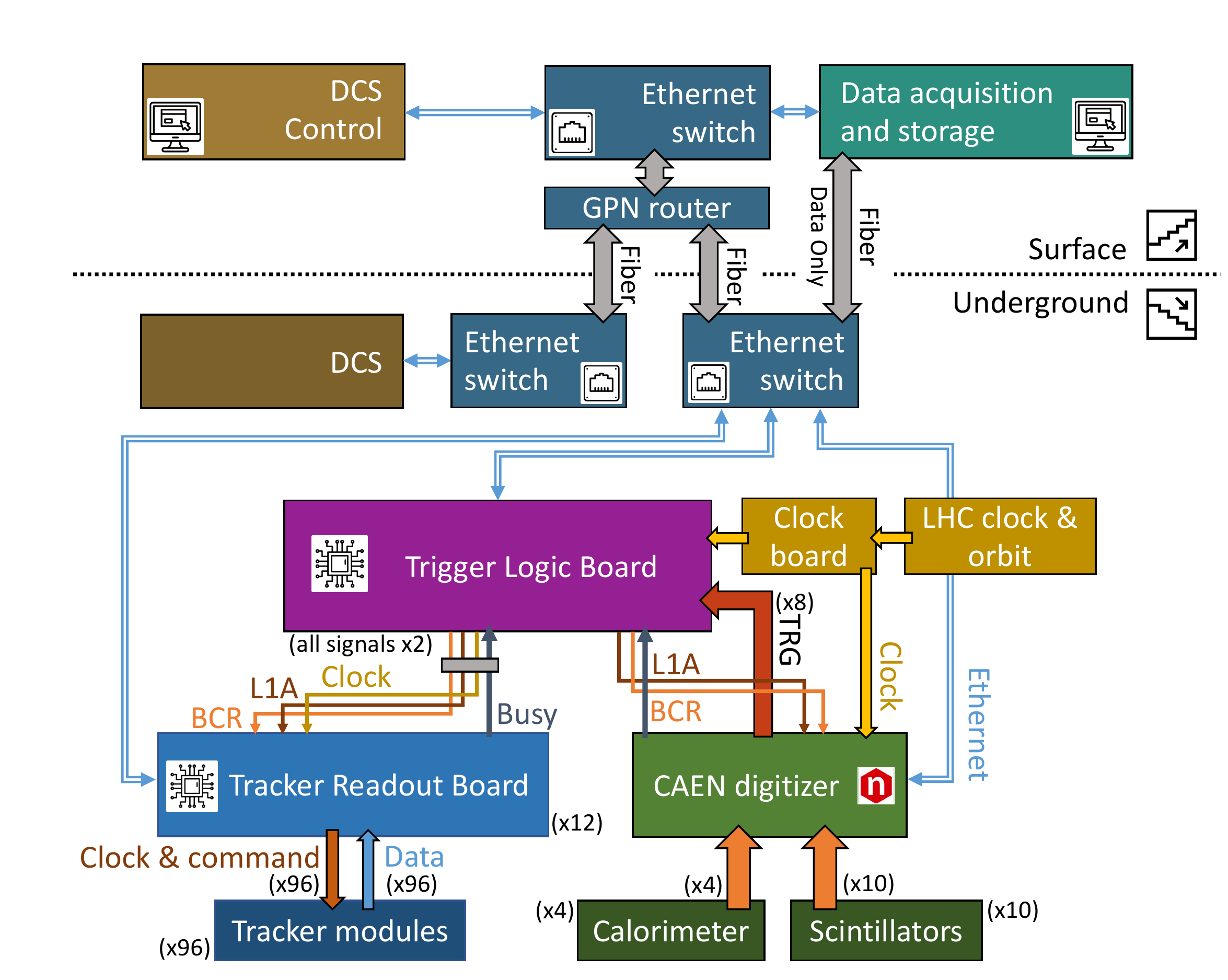}
    \caption{A simple schema of the FASER TDAQ architecture. The numbers in parentheses indicate the number of channels or lines. The blue double-line arrows indicate the connections via Ethernet. The grey thick arrow indicates fibers from the TI12 tunnel to the surface. The digitizer, clock board and TLB are housed in a VME crate. The TRBs are located in two dedicated mini-crates just above the tracker structure, one of the two housing the IFT TRBs.} 
    \label{fig:FASER_TDAQ_schema}
\end{figure}

The FASER TDAQ system is designed to maximize stability and robustness  during data taking, to ensure that the FASER experiment will record with high efficiency data that may contain rare new physics processes. To achieve this, the FASER TDAQ system employs a simple architecture, shown in figure~\ref{fig:FASER_TDAQ_schema}, that minimizes the number of different hardware components, as well as the cables and equipment in the experimental area, which will be inaccessible during data taking. The FASER experiment is planned to be operated remotely only, without shifters populating a control room: the TDAQ system relies on sufficient monitoring to guarantee its robust functioning.

The experiment will be triggering on any high-energy particle traversing its detector volume. It is expected from simulations and in-situ measurements~\cite{FASERTP} that at an instantaneous luminosity of $2\times 10^{34}$~cm$^{-2}$s$^{-1}$ about 500--1000~Hz of particles, dominated by muons originating from  the ATLAS IP, will leave signals in the FASER scintillators, while about 5~Hz of energetic signatures will be deposited at the FASER calorimeter. The digitizer, connected to the scintillators and calorimeter channels, generates trigger signals when the detector signals exceed a preset threshold; the scintillator trigger threshold will be below that of a single minimum ionizing particle, while the calorimeter threshold will be set to trigger on electromagnetic showers depositing more than about 20 GeV of energy. A separate trigger signal arrives from the LED calibration system. The trigger signals are received by the Trigger Logic Board (TLB), the central triggering board of the FASER TDAQ system, which subsequently provides a global trigger accept signal (L1A). The L1A signal is transferred to the detector readout boards (the tracker readout board, TRBs, for the tracker, as well as the digitizer for the scintillators and the calorimeter) for reading out their data. The readout boards receive two additional signals: a clock signal, synchronous to the LHC 40.08~MHz clock, coming from a dedicated LHC board and stabilised in a clock-cleaner board, as well as a bunch counter reset (BCR) signal, generated by the TLB on every LHC orbit signal. The TLB also generates the bunch counter ID (BCID), indicating the number of clock cycles that have passed between the last BCR and the trigger signal. 


The data of the various detector components is read out upon a L1A. Data fragments are transmitted over gigabit Ethernet from the TLB, the TRBs and the digitizer to the Event Building processes running on a conventional server on the surface.
Time-stamp and run number is added and the full event is recorded to disk. Additional data quality monitoring is implemented in other processes running on the same server. The data size is about 25~kBytes per event, dominated by the PMT readout. 

All software components of the TDAQ system 
are implemented on top of an open-source lightweight C++ software framework, DAQling~\cite{daqling}, developed at CERN to be used as the core for DAQ systems of small and medium-sized experiments. DAQling offers configuration management, control of distributed applications, monitoring and messaging functionality, as well as a set of generic utilities. It has been adapted, with the support of its developers, to the needs of the FASER DAQ system.

The FASER detector is controlled and monitored via the DCS, connected to two dedicated servers located on the surface, running the standard LHC Supervisory Control and Data Acquisition (SCADA) system, WinCC OA~\cite{WinCC}. All DCS components in the experimental cavern are connected to a dedicated Ethernet switch, from where data (commands) are transferred to (from) the servers on the surface. Standard industry protocols (OPC-UA / MODBUS-TCP) are used to handle the communication between the different devices controlled and monitored by the DCS and the software back-end. The DCS makes use of commodity devices where possible, such as high and low-voltage power supplies. An electronics board (the TIM board) was developed for FASER and is used for monitoring environmental parameters of the tracker and as a hardware-level interlock to ensure the safety of the tracker.

\section{TDAQ hardware components}\label{sec:trigger_hw}

This section describes the hardware components of the FASER TDAQ system. It starts with the trigger components, specifically those  that feed signals into the TLB: the clock board (section~\ref{sec:clock}) and the digitizer (section~\ref{sec:digi}). It continues with a description of the custom board
(section~\ref{sec:gpio}) used for the implementation of both the TLB and the TRBs. The TLB, representing the main trigger board of the experiment, is discussed in detail (section~\ref{sec:hw_tlb}). A description of the TRB follows (\ref{sec:TRB}). The section closes with presentations of the crates where these boards are housed (\ref{sec:crates}), as well as the DAQ components (\ref{sec:daq_hw}).   



\subsection{LHC signals and the FASER clock}\label{sec:clock}
The LHC clock (40.08\,MHz) and orbit signal (11.245\,kHz) are part of the beam synchronous timing (BST) system, transmitted to beam instrumentation equipment around the LHC over optical fibers using the TTC system\,\cite{TTC}. For FASER, this signal is received over a single optical fiber by a legacy VME system, the BST receiver interface for beam observation system (BOBR)\,\cite{BOBR}, produced by the LHC beam instrumentation group. This board extracts the LHC clock and orbit signals from the optical signals and converts them to electrical signals output on two front panel LEMO connectors. The LHC clock provided by the BOBR has a non-negligible jitter mainly due to noise in its power module, changes during the energy ramp of the LHC, and is not guaranteed to be continuous when there is no beam in the LHC.

The electronic boards of the TDAQ system 
require a high-quality, uninterrupted reference clock that has a constant phase with respect to the LHC clock across power-cycles. The reference clock is provided to FASER by the FASER clock board (FClock), which cleans the jitter of the BOBR  to less than 4~ps, measured with a Time-Interval-Error method.

\begin{figure}
\centering
\includegraphics[width=0.69\textwidth]{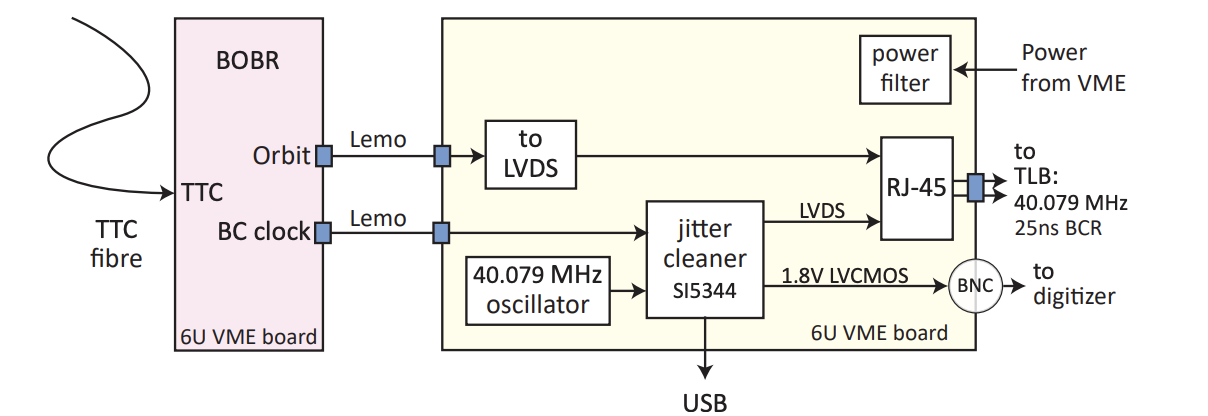}
\caption{Block Diagram of the FASER clock board.}
\label{fig:FClock}
\end{figure}

The FClock is composed of an off-the-shelf clock-cleaner board (Si5342-D-EVB) mounted as a mezzanine on a purpose-specific VME adapter board. The FClock is powered through the VME interface, and the cleaned clock is sent as an LVDS signal via a RJ-45 connector to the TLB. The orbit signal, a 25\,ns pulse centered on the rising edge of the clock, is converted to LVDS and sent on the same RJ-45 connector. 
A single-ended 1.8\,V LVCMOS copy of the clock is also provided on a BNC connector directly to the digitizer. A block diagram of the FClock board is shown in figure~\ref{fig:FClock}.

The zero-delay feature of the jitter cleaner guarantees
that the output FASER clock is aligned to the LHC clock with its phase unchanged across resets and power cycles.
During the LHC cycle the frequency changes up to 550\,Hz and the clock card allows to keep the FASER systems synchronised with the LHC clock.

\subsection{PMT digitizer board}\label{sec:digi}
The readout electronics of the calorimeter and scintillator detectors need to sample the signals from the calorimeter and scintillator PMTs, provide trigger input signals to the TLB and buffer the calorimeter and scintillator data, which are read out upon a trigger L1A. 
This functionality is accomplished by using a VME-based system consisting of a single 16-channel, 14-bit CAEN VX1730 digitizer board~\cite{VX1730}  that is controlled and read out via Ethernet using the Struck Innovative Systems SIS3153 VME interface board~\cite{SIS3153}. The dynamic range of a digitizer channel is 2~V with a programmable offset of up to $\pm1$ V that can be set independently on each channel. The signals on each channel are continuously digitized at 500~MHz and stored in a circular buffer to be read out on receiving a trigger signal. The on-board trigger logic runs at 125~MHz and for each channel a user-defined, fixed, absolute threshold can be set to generate an over/under-threshold signal. These are combined in pairs into eight trigger signals using a logical AND or OR, depending on the need. The eight trigger signals can either be used to trigger the readout internally for standalone testing or transmitted to the TLB as LVDS signals for logical combinations. In the latter case, the TLB L1A trigger decision is transmitted over a TTL LEMO signal from the TLB to the digitizer.

Upon receiving a L1A trigger signal, the board transfers the recorded samples into a FIFO readout buffer. The recorded data contain the data from all of the enabled channels as well as a 4-word data header. The data header contains the event counter and the time since the last BCR received from the TLB, both of which are used to identify the event. The time is measured in units of 16\,ns which is later converted into the BCID in software. The full waveform for each channel is read out for a user-defined period, which in FASER is foreseen to be about 1.2 $\mu$s at the beginning. This allows for a detailed offline analysis of the waveform, in particular in the case of any anomalous signal, at the cost of a rather large data size. This can later be reduced by a factor three (limited by the trigger latency) if no anomalous behavior is observed in early data. More advanced firmware with zero-suppression or digital pulse-processing are available from CAEN, which could reduce the data size further. These are currently not foreseen to be used. The readout FIFO is large (up to 1024 events), but can get full in case the readout of the board is unable to keep up. To avoid overflows, the board will raise a `buffer-full' signal over the LVDS connection to the TLB once the buffer occupancy reaches a pre-defined threshold, which will halt any further triggers until occupancy drops below threshold.

The VX1730 board digitizes the signals at a frequency of 500~MHz. Its internal logic runs at 125~MHz while the readout is happening at 62.5~MHz. These frequencies cannot be synchronised to the LHC clock of 40.08~MHz. This leads to an intrinsic jitter for the trigger signals of 8~ns. 
To allow for precise measurements of the arrival time of each signal with respect to the collisions in the LHC, a copy of the LHC clock supplied from the FClock card is sampled on one digitizer channel. Offline data analysis uses a Fast-Fourier Transformation of this digitized clock to measure the phase of each event very precisely.

The configuration of the digitizer and the readout of the data FIFO buffer is performed using the Ethernet interface of the SIS3153 VME board. It communicates with the digitizer board using the VME backplane and translates the VME commands and data to simple UDP Ethernet packets, thus avoiding the need for a dedicated single-board computer in the VME crate.

\subsection{General Purpose I/O module}
\label{sec:gpio}

\begin{figure}
  		     \centering
  		     \includegraphics[width=0.69\textwidth]{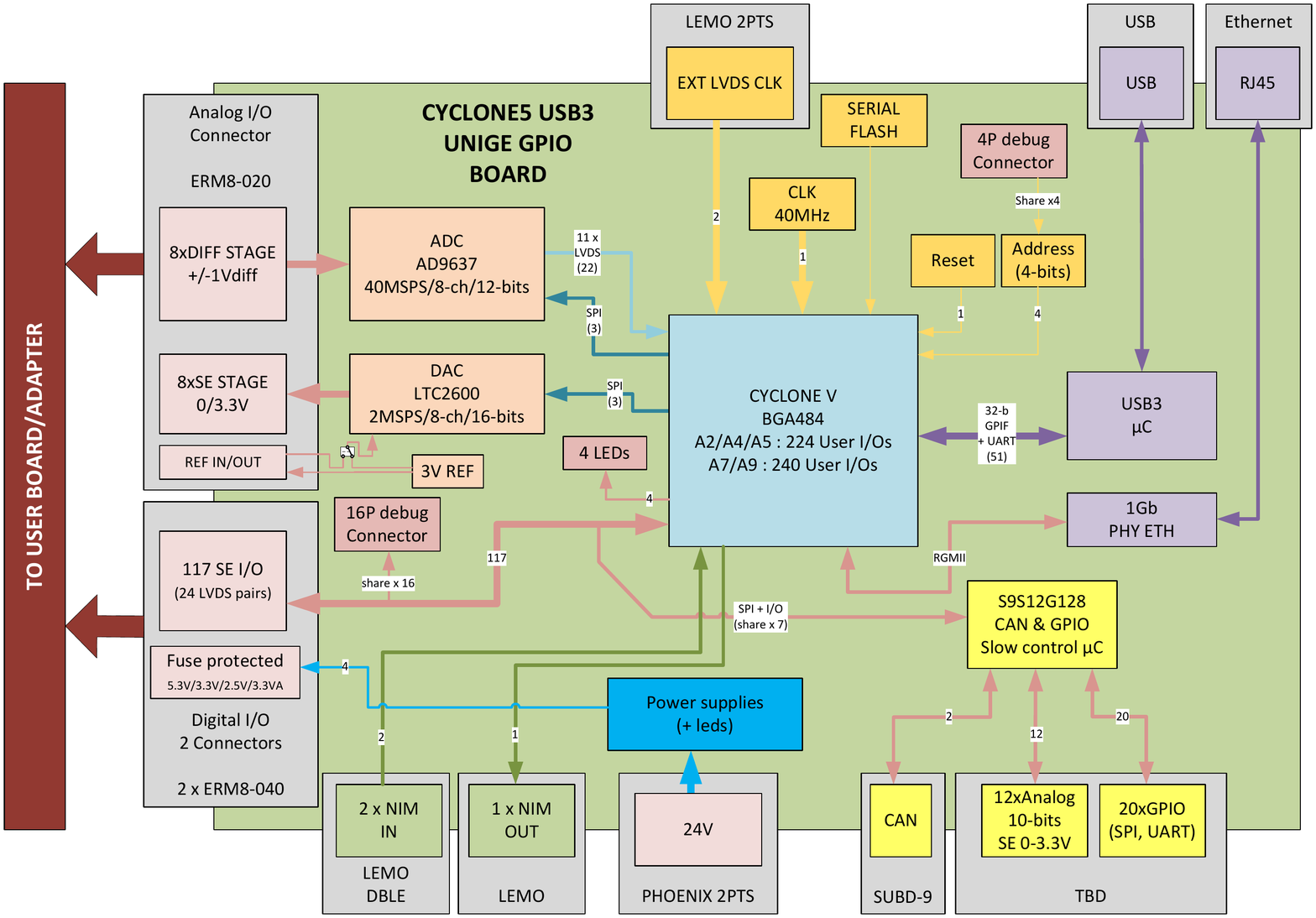}
  		     \includegraphics[width=0.29\textwidth]{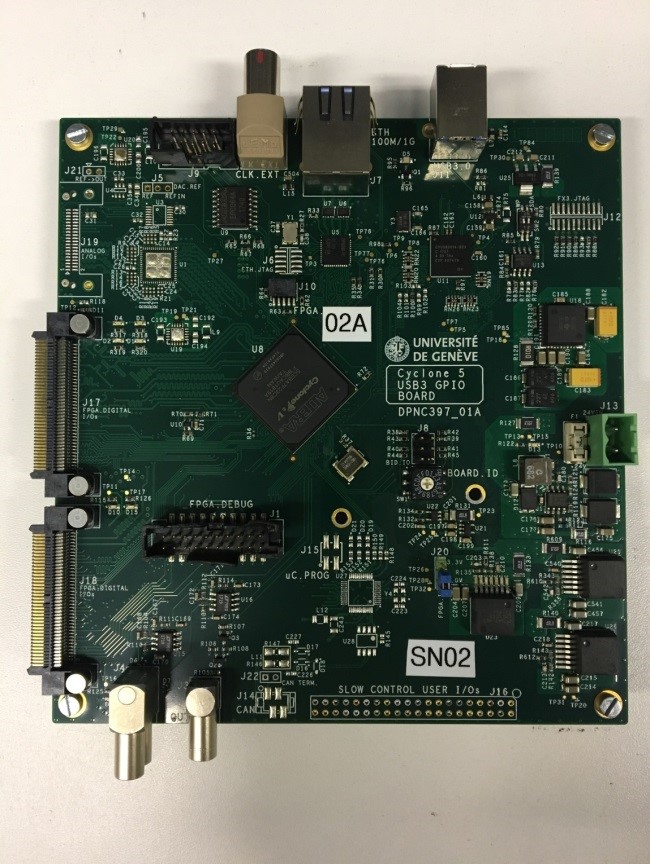}
  		     \caption{The architecture of the ``Unige GPIO" board and a picture of the assembled board.}
  		     \label{fig:GPIO}
  		 \end{figure}

The TLB and TRB boards are both based on the `Unige GPIO', a general purpose board developed by the University of Geneva for slow control and readout of particle physics ASICs, detectors tests, and qualification. For each application, such as for the TLB and TRB in the FASER experiment, a dedicated interface board needs to be developed to interface to the hardware. This interface board provides a connection for the analogue and/or digital inputs/outputs. 
The architecture of the `Unige GPIO' board and a picture of the assembled board are shown in figure~\ref{fig:GPIO}.
  		 
The core of the board is composed of a CYCLONE V A7 FPGA with 56.5K adaptive logic modules, ALMs, with 8 input look-up tables, LUTs, and 4 registers per ALM, 6.86~Mbits and 836~kbits of RAM per input type, respectively, 156 double 18x18 digital signal processor, DSP, blocks and 7 phase-locked loops, PLLs. The FPGA is driven either by a 40~MHz local clock or by an external clock via LVDS input through a dedicated 2 pin LEMO connector.
For the board to interface with the experiment, 
two connectors dispatch 117 single-ended inputs/outputs, which can be configured through the FPGA firmware. 
More signals are available in the board but not used in FASER, neither in the TLB nor in the TRB.

Two interfaces can be used for readout and slow control. The first one is USB3, where the USB protocol is fully handled by an on-board FX3 Cypress micro-controller providing a 32-bit parallel
interface clocked at 100~MHz. The FPGA has a maximum potential USB data throughput of 3.2~Gbit/s,
but is ultimately limited to 1.8 to 2.5~Gbit/s depending on the host PC performance when connected with the
board. The second interface is a 1~Gbit/s Ethernet with an on-board hardware chip driving the PHY layer, an
Intel IP implemented in the FPGA for the MAC layer, and a 50~MHz Intel NIOS2 soft microcontroller
running the ARP, IP, ICMP and UDP protocols used by the readout and the slow-control. 
On top of
the two communication interfaces, a full and versatile FPGA library is  available for the readout and the slow
control of any small physics experiment or test bench. The firmware user
interface is the same for both USB and Ethernet allowing the possibility to switch easily from one to
another with just a synthesis of the firmware design.
The communication is based on 32-bit words that encode a command word identifier, a board identification number of the board that is communicated with, the command type, and 
up to 16-bits of payload. The board always answers with similarly structured 32-bit words, acknowledging all known commands
or returning an error in case any part of the command cannot be interpreted or the data the command requests contains an error. The board uses two independent data streams. The first is for bi-directional serial communication
used to send commands and send or read back configuration data. The second is a uni-directional stream from the board to the host-PC transferring the 
event data to be recorded. 

One single 24~V power supply input is required to power the board and 5.0~V, 3.3~V and 2.5~V are accessible through the user connectors for the interface board.

\subsection{Trigger logic board}\label{sec:hw_tlb}

\begin{figure}[h]
    \centering
    \includegraphics[trim=1cm 2.5cm 4.5cm 4cm, clip=true, width=\textwidth]{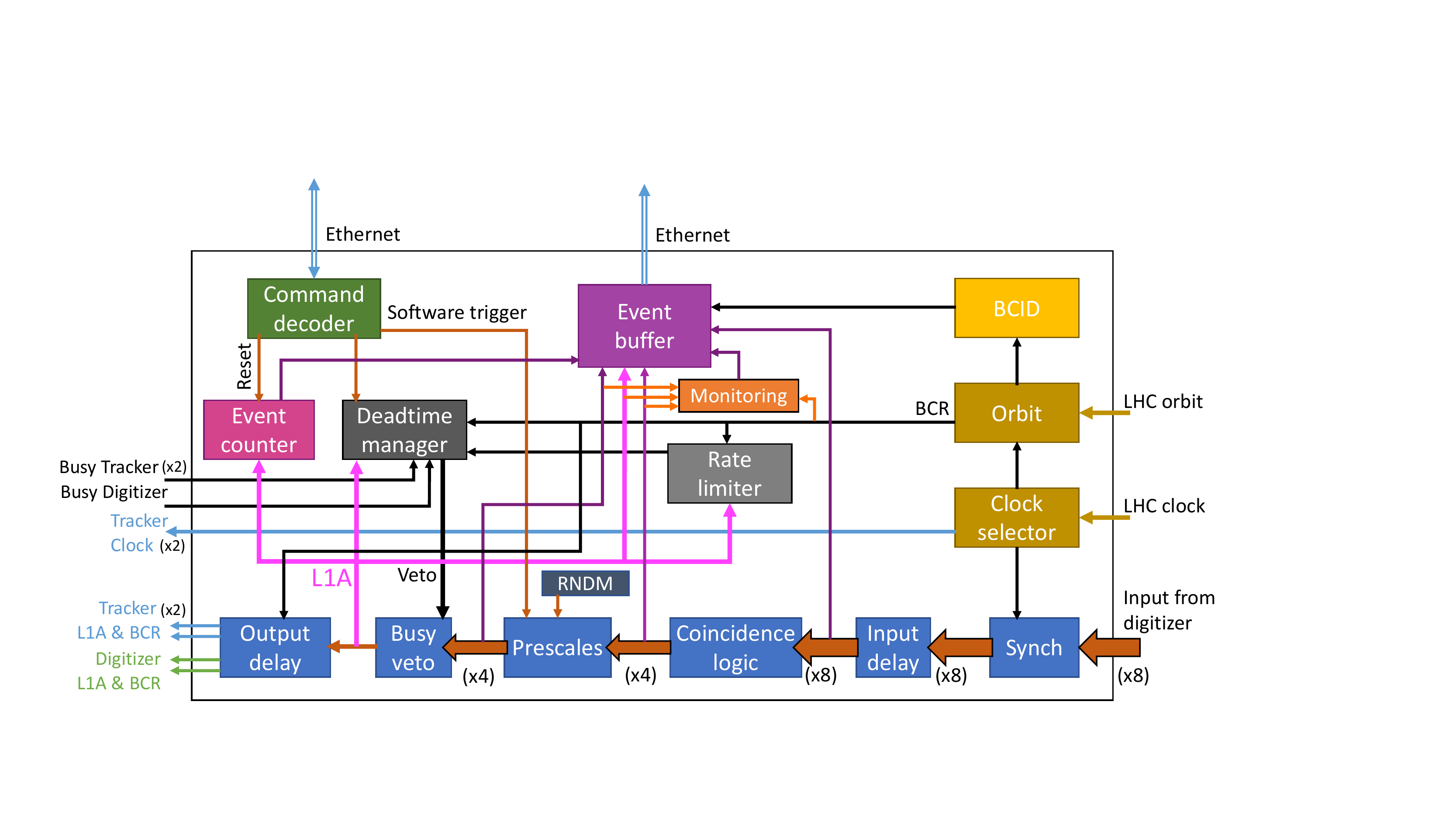}
    \caption{Schematic diagram of the functionality implemented in the Trigger Logic Board.} 
    \label{fig:FASER_TLB_schema}
\end{figure}

The TLB is the central trigger logic processor and distributor of a common (LHC or internal) 40.08~MHz clock and bunch-crossing reset signal. It receives and combines up to 8 trigger input signals from the digitizer and transmits the final trigger decision to all readout components. In addition it manages the output rates via prescaling and veto functionality. The functionality implemented in the TLB is schematically shown in figure~\ref{fig:FASER_TLB_schema}.

The TLB functionality is implemented in the firmware of a GPIO board described in section~\ref{sec:gpio}. 
It receives the LHC clock signal and BCR LVDS signal via a RJ45 connection. The BCR signal is distributed to the digitizer as a transistor-transistor logic, TLL, signal over a LEMO connection, while both the clock and BCR signal are transmitted as LVDS signals over two 15~m CAT7 cables to the crates that house the tracker readout boards.  The TLB receives the trigger signals from the digitizer via a short, flat LVDS cable, and the trigger signal from the LED calibration system via a separate TTL signal on a LEMO input directly. Finally, the board receives and replies to slow control commands and transmits its data via Ethernet\footnote{While Ethernet is used in the final system, the board fully supports \texttt{USB3} communication as well, as described in section~\ref{sec:gpio}.}.
Both the digitizer and tracker can assert a busy signal.

The TLB synchronises the digitizer trigger signals to form 8 trigger lines in total. The 8 trigger lines (8 bits) are combined into 4 physics trigger items (4 bits) based on a \LUT{} with a predefined coincidence logic loaded at configuration time. Two additional trigger  bits are defined: an internally generated pseudo-random or fixed rate trigger, and another for the LED calibration signal but which can also be activated by a software command.
Each item can be `prescaled' by a factor $n$, meaning that only every $n$th signal is sent further. Finally, a logical OR between all trigger items is computed and defines the final trigger decision - or L1A - which is then transmitted to all readout components unless vetoed; several sources, further described below, can create a non-maskable veto.
Upon an L1A, the TLB will send out a data packet containing: the event ID, BCID, orbit ID, trigger items before and after prescale, and the bits of the 8 trigger lines for the triggered and the following \BC{}.
In addition, the TLB regularly publishes monitoring data packets containing trigger item counters before and after prescale (TBP and TAP) and after veto (TAV), veto counters, and the logical OR of all trigger items before and after veto to enable the computation of the total fraction of vetoed events.

\begin{figure}[ht]
\begin{center}
 \includegraphics[width=\linewidth]{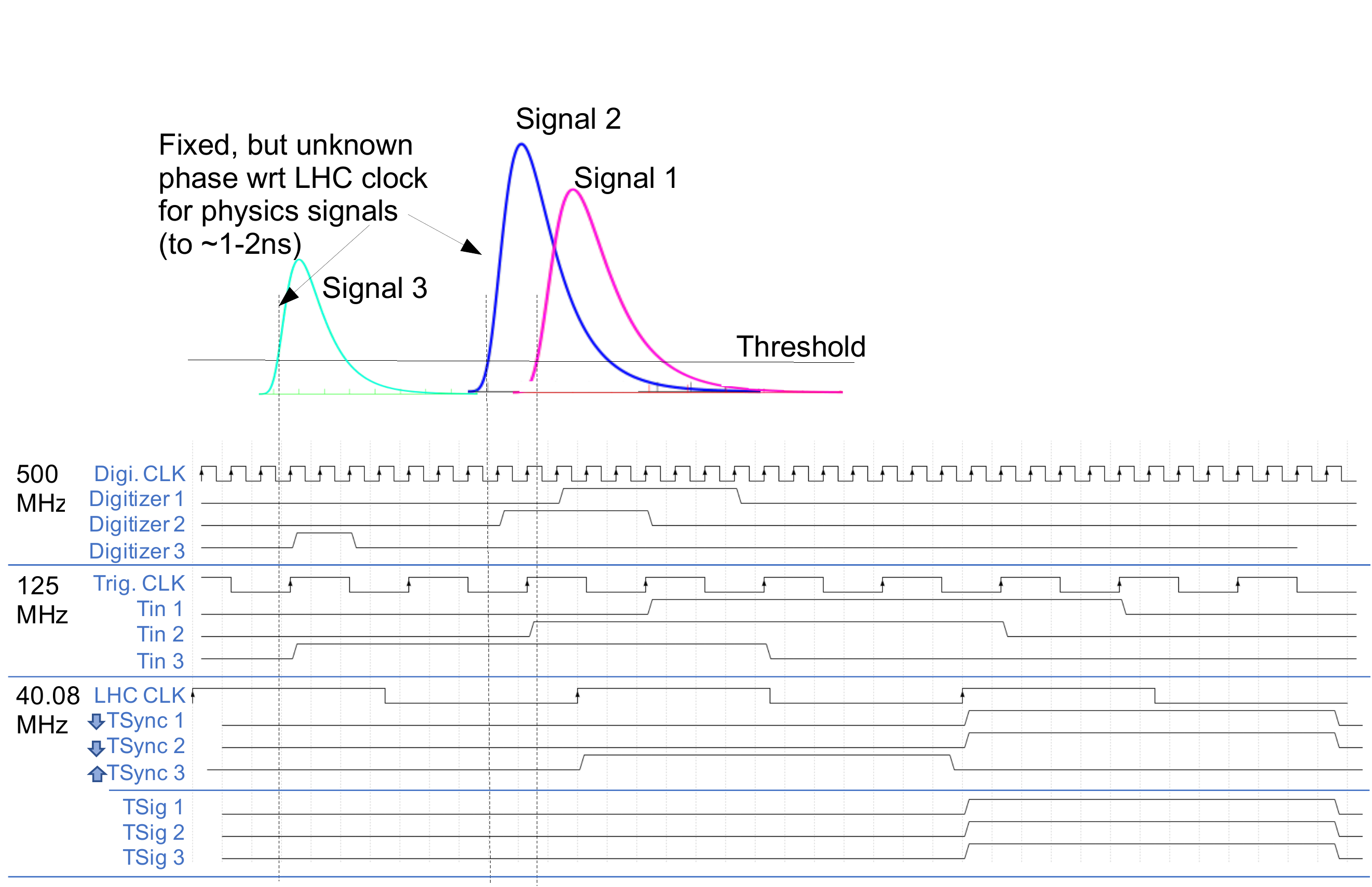}
 \caption{Example of input signals and how signals are synchronized. Only 3 of the 8 input channels are shown. The digizer runs at a 500~MHz clock. The digitized input signals (Digitizer 1, 2 and 3) are sampled at 125~MHz (Tin 1, 2 and 3) and with a 8~ns jitter to produce the trigger signals. These signals are synchronised with the LHC clock using either its rising or its falling edge; this option is configurable and is tuned in dedicated runs before data taking. In this example, the first two input channels are sampled for synchronization on the falling LHC clock (LHC CLK) edge, while the third channel is sampled on the rising edge as indicated on left in the figure. The synchronised signals (TSync 1, 2 and 3) can then be delayed so that the final trigger signals (TSig 1, 2 and 3) are all synchronised with each other; this option is also configurable and tuned in dedicated runs prior to data taking. In this example, TSig 3 is delayed by one clock cycle.}
 \label{InputTiming}
\end{center}
\end{figure}

\paragraph{Trigger timing}
It is important that the input signals from the digitizer can be correctly synchronised to the LHC clock. The trigger input signals from an event may not arrive simultaneously because of differences in the time-of-flight of particles traversing the length of the detector and the different PMT response times. 
Figure~\ref{InputTiming} shows a timing plot of 3 incoming signals and their subsequent synchronisation.
The input (Tin) signals appear 32 ns wide, thus covering over a full LHC clock cycle. The 8 ns jitter is avoided by having the option of sampling each input signal on the rising or falling LHC clock edge, as it is guaranteed that at least one edge is in phase with a stable section of the input signal; this option will be tuned in data taking with beam. 
The difference in arrival times between input signals may be large enough that the output (TSync) signals synchronised to the LHC clock do not land in the same clock cycle. An input delay can therefore be applied to each channel to synchronise the final trigger line output (TSig) signals.
Once an L1A is generated, the L1A signal sent to the digitizer and the tracker can be separately delayed with a granularity of one \BC{}. This is to ensure that the readout window on each readout board is aligned with the complete data corresponding to the event that triggered the L1A.
The final adjustable delay is the orbit delay. This delays the BCR transmission so that it can be tuned to occur during the LHC abort gap when no proton collisions occur.

\paragraph{Trigger veto}
Different sources for trigger vetoes ultimately serve to limit the minimum spacing between consecutive triggers when subdetectors are busy reading out event data, or to limit the overall trigger rate, leading to system \textit{deadtime}. The possible sources of trigger vetoes are listed below. The TLB will veto events while receiving a busy signal from a subsystem.

\begin{itemize}
\item \textit{Tracker busy veto}: The TRBs can only receive and readout one L1A at a time. Each tracker readout board will assert a busy signal while its readout FIFOs 
are being read (see section~\ref{sec:TRB}). The duration of the busy signal is thus dependent on the hit multiplicity in an event.
\item \textit{Digitizer busy veto}: The digitizer will assert a busy signal if its internal readout buffer occupancy reaches a certain threshold. The busy signal is held until its buffer occupancy drops below threshold.
\item \textit{Simple deadtime veto}: All trigger decisions within a chosen fixed number of \BCs{} after a L1A are vetoed. The simple deadtime is at minimum equal to the L1A delay to the tracker readout to avoid triggers before the tracker busy signal sets in. The combined simple plus tracker busy deadtime is expected to be on average 300 BCs in duration.
\item \textit{BCR veto}: Since SCT modules require 3 clock cycles to process a L1A and 7 clock cycles to process a BCR, an overlap of these is prevented by vetoing triggers for 9 clock cycles starting from 2 clock cycles before a BCR. This implies a fixed deadtime of 0.25\% for random or cosmic-ray events, but will be zero for collision events as the BCR will be tuned to occur during the LHC abort gap.
\item \textit{Rate limiter}: In order to avoid uncontrolled high rates from an input channel suddenly becoming noisy, a rate limiter limits the L1A rate to 3 L1As per 15 orbits (or 2.2 kHz).
\end{itemize}

 The dominant sources of deadtime are expected to be the combination of simple and tracker busy deadtime of 0.4\% (0.7\%) at 500~(1000) Hz, and the rate limiter, which would start dominating at 1~kHz at 1\% deadtime. 
 The BCR veto is expected to induce no deadtime for physics events as the BCR will be tuned to occur during the $\mathcal{O}(\mu s)$ LHC abort gap, while digitizer deadtime is expected to not be a limitation during stable running. A final measurement of the deadtime with respect to the expectation for a range in trigger input rates is presented in section~\ref{sec:combined_com} (see figure~\ref{fig:ehn1_rate_vs_deadtime}).

\paragraph{TLB internal trigger generation}
The TLB can generate an internal trigger in 3 different modes: (i) in fixed rate mode (ii) in pseudo-random rate mode and (iii) a single trigger via a command (software trigger).
In fixed rate mode, a flat rate of a maximum of 100~kHz can be set. 
In pseudo-random rate mode, a total of 8 settings allow for an unprescaled average range in rate of roughly 15~Hz to 1.2~kHz. The pseudo-random trigger relies on a 22-bit seed, which is updated at every clock cycle via a linear-feedback shift register. The seed is then aligned with a 22-bit mask that has the 13+N leading bits turned on, where N is the random trigger setting of 0--7. A trigger is fired if the seed and the AND of the mask align to zero. For instance, a random trigger setting of 0 (7), means a trigger will fire when the first 13 (20) bits of the seed are zero. A higher random rate setting therefore leads to a less frequently fired trigger as a greater number of zeros are required to align.
The seed is incremented by one on every L1A to avoid repeating patterns in the linear-feedback shift register.


\subsection{Tracker read-out}\label{sec:hw_trb}
\label{sec:TRB}
\label{sec:hw_trb}

The tracker readout board, or TRB, handles all communication with the attached tracker modules.
It is responsible for configuration, propagation of trigger signals and readout of the data. It is based on the GPIO board described in section \ref{sec:gpio}.
The TRB itself receives trigger (L1A) and timing (clock and BCR) signals from the TLB, described in \ref{sec:hw_tlb}.
Up to 8 SCT modules can be connected to one TRB via an adaptor board. The TRBs are located in a mini-crate, described in section~\ref{sec:crates}.
The adaptor board interfaces with the backplane of the mini-crate, which provides the timing and trigger signals. 
The board does not decode the module data except for finding the end of data marker. Its functionality can be divided into three parts:  
Communication with the host PC, configuration of the tracker modules and data taking. 

In order to configure the tracker modules the TRB must send configuration commands and data 
encoded in a bit-stream to the tracker modules. To this end, the TRB firmware implements functionality to prepare the bit-stream from 
commands and data received from the host PC. 
An arbitrary combination of modules can be addressed simultaneously. 
In data taking mode the TRB transmits the trigger and timing signals to the enabled modules. 
When in this mode, two running modes are distinguished: calibration and physics data taking.

During physics data taking the L1A, BCR and clock signal are received from the TLB, processed in the TRB and sent to the modules
with a fixed latency. Since the tracker readout has an internal pipeline of 132 \BCs{}, the TLB L1A signal must arrive at the SCT modules 132 BCs after the charged particle traversed the tracker plane. There are inherent delays in signal propagation from the occurance of an event to the arrival of the L1A at the TRB, but to cover the full 132 BCs latency, both the TRB and TLB have a settable coarse time delay in BC increments for the L1A/BCR signals. A signal delay of up to 127 BCs on the TLB and a further adjustment of 0 to 7 BCs on a TRB are possible, where the latter allows to adjust for different particle times-of-flight at each tracker station location in the final detector setup. The tracker readout timing can be further adjusted with a resolution of 390~ps via an adjustment of the input clock phase on the TRB, so that hits are sampled on the pulse peak.
The data received from each module upon an L1A are stored in the TRB in so-called Level 1 FIFOs - 16 FIFOs in total, as there is one per each side of a module.
When any of these FIFOs are filled with data, the TRB asserts a hardware busy signal that is sent over the
backplane to the TLB.
Each event is associated with a BCID, calculated within the TRB using an internal counter and the BCR signal, and which is used later in the processing for data consistency checks, as described in section~\ref{sec:daqsoftware}. 

In calibration mode the L1A and clock signal are generated in the TRB, the former upon a command received from the host PC. 
It is possible to send one L1A at a time. However, every command sent to the TRB by software requires sending at least one 32-bit word and waiting for a reply from the TRB. This implies that the rate of sending individual trigger commands is given by 
the latency of the data transfer of single words. This is in the order of milliseconds 
for UDP and tens to hundreds of micro seconds for USB. Thus the trigger rate is limited to a few hundred Hz when 
using the default UDP interface, which is unacceptably slow for calibration scans that require millions of triggers. 
To overcome this limitation, a firmware feature allows sending a burst of up to 4095 triggers with 
an adjustable latency in between. This allows trigger rates that are only limited by the speed with which 
the data is shipped to the readout PC. To avoid loss of data the delay between consecutive triggers is 
adjusted dynamically depending on the remaining space in the receiving FIFO.
Complete events are formed and stored in a Level 2 FIFO until pushed out. 
A complete event consists of a header, containing the event ID.
The header is followed by one or several TRB data words, containing the BCID as well as any errors that might have occurred. 
Appended to this is the module raw data 
prefixed with the module ID.  
Every word holding part of the event data contains a frame ID as well as a consecutive data frame counter to allow for reconstruction of the full
event information. Additionally a cyclic redundancy check (CRC) value is computed over the full event data and can be used to identify any transmission errors.

\subsection{Crates}\label{sec:crates}

\begin{figure}[h]
    \centering
     \includegraphics[width=0.56\textwidth]{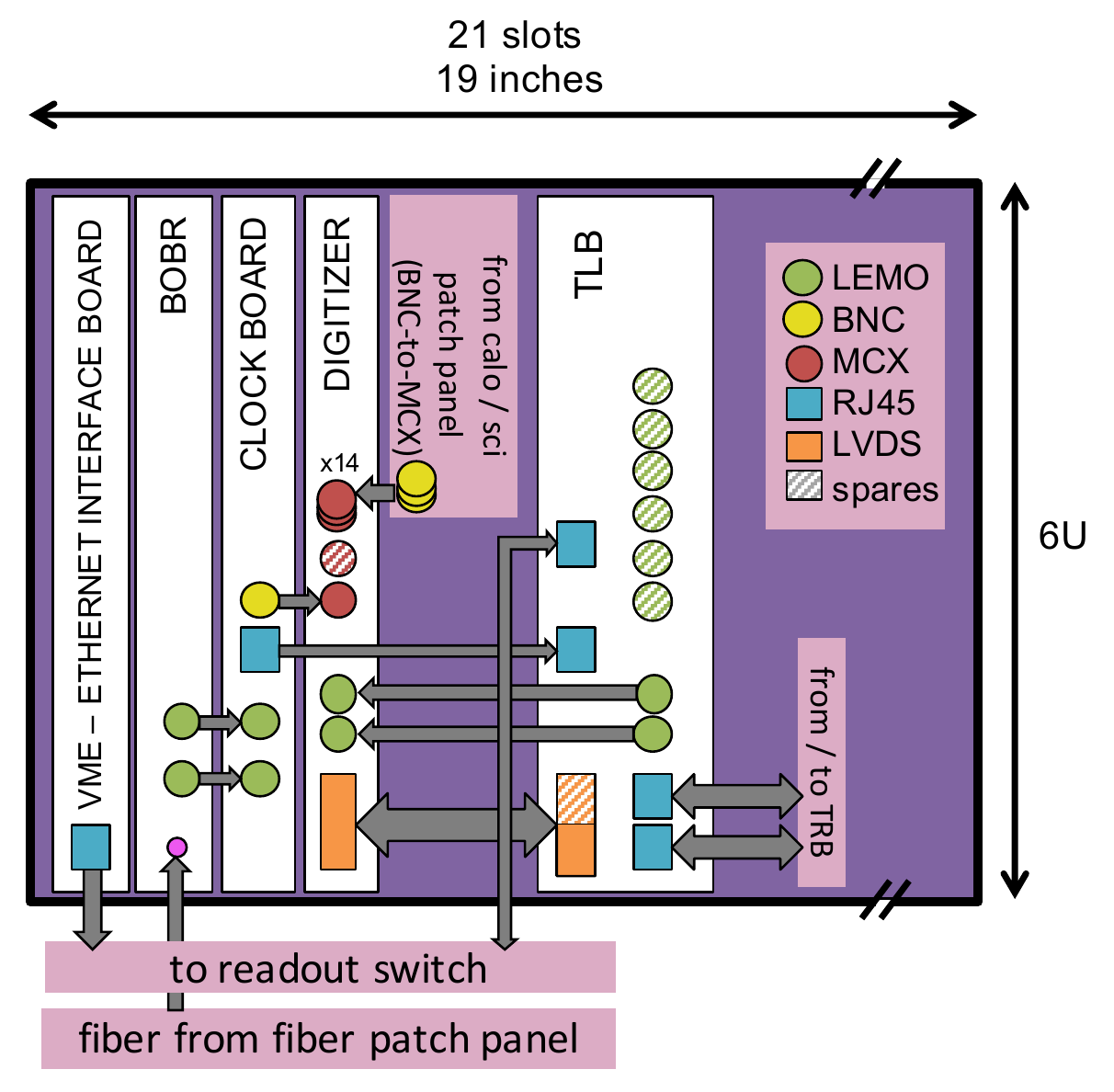}   
     \includegraphics[width=0.41\textwidth]{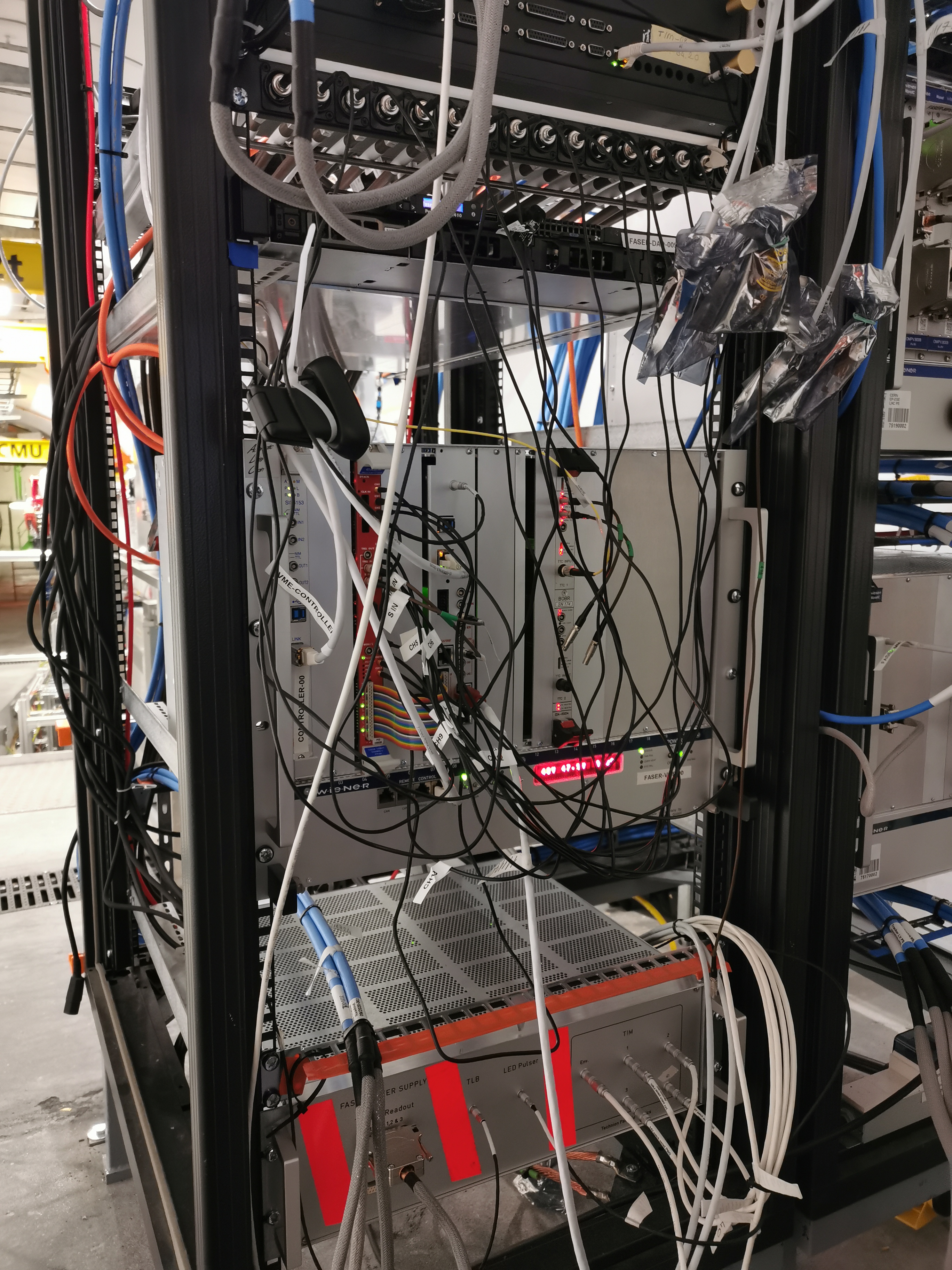}
    \caption{The VME crate housing the BOBR, clock board, TLB and digitizer: a schematic view of the connections (left) and a picture of the rack where the crate is located in TI12 (right).} 
    \label{fig:VME}
\end{figure}

The FASER trigger components are housed in two crates. The TRB modules use a dedicated mini-crate that is located in a support  structure just above the middle tracking station. 
The backplane is composed of 9 slots allowing the connection of 9 TRBs\footnote{The 3 TRBs used for the IFT station will be housed in an identical mini crate situated on the detector close to the IFT.}. Individual LVDS pairs transfer CLK, L1A and BCR signals, and an open drain output transfers the BUSY signal from the TRBs. One temperature sensor and the 24~V power per TRB are also transferred through these slot connectors.
Local DC/DC converters are also located on the backplane and allow the 3.3~V powering of 3 fanout chips used for the distribution of the individual TRB LVDS BCR, CLK and L1A from the single RJ45 connector connected to the TLB.
The rest of the trigger boards are inserted in a standard VME crate\footnote{The model of the VME crate is the following: \url{http://www.wiener-d.com/sc/powered-crates/vme64-x/6u-vme64x-6021.html}}. A schema of the front panels of the boards and how they are connected are both shown in figure~\ref{fig:VME}, alongside a picture of the crate, located in the rack in TI12.

\subsection{Computing and networking components}\label{sec:daq_hw}

The readout of the FASER front-end electronics as well as the detector control system is based on commercial Gigabit Ethernet using both copper and fiber-optic connections. As illustrated in figure~\ref{fig:FASER_TDAQ_schema}, all Ethernet connections from the electronics cards and crates in the TI12 tunnel are aggregated in two Ethernet switches in TI12 for the DAQ and DCS signals, respectively. The DAQ switch is connected over more than 1.5\,km long single-mode 10~Gbit/s optical fibers directly to a single DAQ server PC in a surface building at the LHC Point-1 area (SR1), while the DCS switch is connected to the DCS server PCs through the main SR1 gateway and a local switch over 1~Gbit/s optical fibers. The DAQ server is connected to the CERN general purpose network (GPN) through the same switch in SR1.
Control of the DAQ and DCS systems as well as the transfer of data files to permanent storage in the CERN computing center are all done through this connection.


The network switches used by FASER are all 24-port Ruckus ICX-7450 switches with one or two additional fiber optics up-link modules as well as redundant power-supplies. The switches are configured and monitored by CERN central IT support.

Tests have shown that a single DAQ server is sufficient to run the full FASER DAQ readout software as described in section \ref{sec:daqsoftware} provided it has sufficient computing capacity. The final DAQ server is a 1U, SuperMicro server with a 24-core, 48 thread AMD EPYC 7402P CPU, 64 GByte ECC DDR4 memory.
Two 480~GB SSDs in RAID 1 are used to host the operating system, DAQ software and local log files, while two 8TB 7200 RPM hard-drives in RAID 1 are used as local disk storage, providing sufficient storage for more than one week of continuous data taking at the nominal trigger rate and data size. The server has redundant power supplies, connected to a central diesel-backed UPS for maximum reliability\footnote{Data taking would be stopped in case of a power failure in TI12 as most of the electronics are not on UPS power.}. An identical, spare server is maintained in the same rack in SR1 as a backup in case of a major failure of the central DAQ server.

The DCS server resource requirements are more modest and two 1U, SuperMicro servers with a 4-core, 8 thread Intel Xeon E2244G CPUs and 16 GByte ECC DDR4 memory are used. Two 480~GB SSDs in RAID 1 are used to host the operating system, DCS software and local log files. As with the DAQ server, the DCS servers have redundant power supplies and a spare server is maintained as backup. All DAQ and DCS servers are running the CERN standard CentOS7 Linux distribution.


For the tests described in this document, the direct connection for the data path from TI12 to the SR1 server was not yet ready. Instead, the connection is done through the main SR1 gateway and the second switch in SR1 over 1~Gbit/s optical fibers. Additional tests are done with a server located in TI12, connected directly to the DAQ switch in TI12.

The final DAQ server was only available for some of the tests described in this document, reported in sections~\ref{sec:DAQ_com} and \ref{sec:combined_com}. For earlier commissioning tests, an older server PC, borrowed from the ATLAS experiment, is used. The
server is a 2012 Dell PowerEdge R410 with two Intel Xeon E5645 processors, each with six cores (twelve threads) running at up to 2.4 GHz. The server is equipped with 24 GBytes of DDR3 memory and a single 500 GByte hard-drive for storage.

\section{DAQ software}\label{sec:software}
\label{sec:daqsoftware}

The software used for the FASER DAQ processes is presented in this section. The FASER DAQ software is implemented on top of an external framework; this implementation is described in section~\ref{sec:daq_sw}. The data format and size are outlined in section~\ref{sec:data_format}. The specific software used for reading out the digitizer and GPIOs (TLB and TRB) is described in sections~\ref{sec:daq_digi} and \ref{sec:daq_gpiodrivers}. Controls and monitoring software are discussed in the following section, section~\ref{sec:controls_monitoring}.

\subsection{DAQ software framework}\label{sec:daq_sw}

The FASER DAQ software~\cite{faser-daq-sw} is built on top of the DAQling framework~\cite{daqling-sw}. 
DAQling is an open-source lightweight C++ software framework that can be used as the core of data acquisition systems of small and medium-sized experiments. It provides a modular system in which custom applications can be plugged-in. It also provides the communication layer based on the widespread ZeroMQ messaging library \cite{ZeroMQ}, error logging using ERS \cite{ERS}, configuration management based on the JSON format, control of distributed applications and extendable operational monitoring with web-based visualization.

\begin{figure}[h]
    \centering
    \includegraphics[trim=1cm 0.5cm 2cm 1cm, clip=true, width=\textwidth]{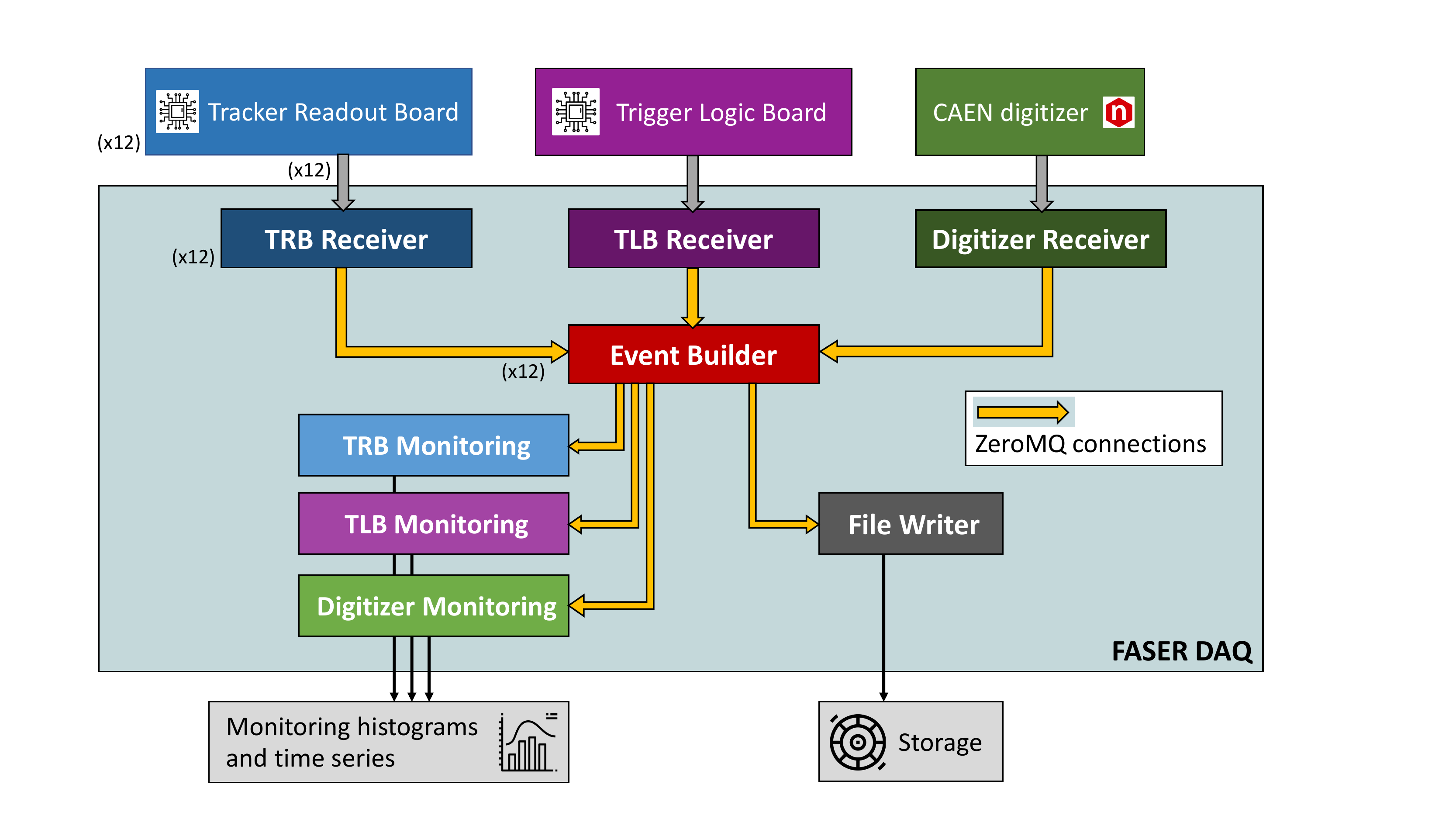}
    \caption{Overview of the FASER DAQ software processes.  }
    \label{fig:DAQSoftwareDiagram}
\end{figure}

The ensemble of FASER DAQ processes and their communication channels are illustrated in figure~\ref{fig:DAQSoftwareDiagram}. Each readout electronics board in FASER is read out, configured and controlled by its own software process, referred to as a receiver process, the details of which are described in the following sections. During data taking, the receivers will assemble the data for each triggered event into the data fragment described in section~\ref{sec:dataformat}, and push these one-by-one to the event builder process. The event builder merges all data fragments for a given event and pushes the full event to the file writer process, which is responsible for recording the events to the local disk. The DAQ processes can run on separate host PCs with communication over TCP with ZeroMQ between them, but since the total CPU requirement is modest, most tests and the final DAQ foresees running all processes on a single host, primarily using ZeroMQ IPC as the communication layer.

The merging of event fragments in the event builder into full events is done based
on the event identifier number assigned in each front-end board, where the boards count the number of triggers they receive and assign the number sequentially. In order to ensure that the fragments truly belong to the same collision, the BCID, stored in the header of the detector data fragments is retrieved and checked.
If the event builder detects an event where the BCIDs for the different fragments do not match, it will flag the event and raise an alarm once a certain threshold of events is reached. The event builder will also flag events where one or more fragments are not received within a programmable time-out and route such events to a separate data stream. The same happens for events where a fragment has been flagged as corrupted by the front-end receiver. The monitoring fragments from the TLB receiver will be routed to a separate stream. The full events in each stream are immediately pushed to the file writer and separately to any subscribing monitoring process, as explained later in section~\ref{sec:monitoring}. The file writer does not decode the received data, but instead just concatenates the events into separate buffers for each stream, writing the buffer to disk once a threshold, typically 1~MByte, is exceeded. Each data stream has its own file on disk, indexed by stream name, run number and a rolling index per run. The files are closed and a new one opened when the filesize exceeds 1~GByte in order to keep the individual data-file at an easily manageable size.

\subsection{Data format and event size}\label{sec:data_format}
\label{sec:dataformat}

\begin{figure}[H]
    \centering
    \includegraphics[width=0.7\textwidth]{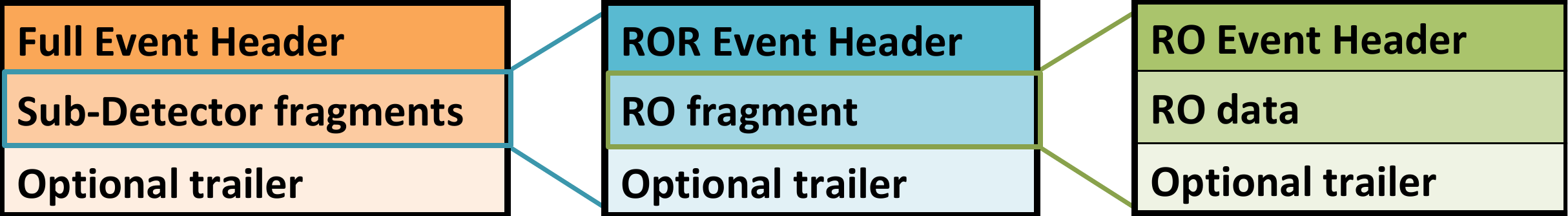}

    \caption{The FASER event format. Sub-detector fragments refer to the TLB, digitizer and tracker. The full event and readout-receiver (ROR) event headers are generic; the readout (RO) event header is sub-detector specific.}
    \label{fig:FASER_eventformat}
\end{figure}

The FASER data format defines the format of the data out of the TLB, digitizer and tracker readout boards, the readout receiver applications, and the data out of the event builder. The event format  provides information redundancy to allow self-consistency checks of the event to be made. The basic unit of the FASER data, after it has been processed by the readout receiver application, is the data fragment. All fragments  have identical structure and the same format of header. The header information  is small in size and the identification of the fragment origin is possible with the header information only. The data format (header and trailer)  provides means of identifying errors related to hardware or data flow issues. A schematic view of the event format structure is shown in figure~\ref{fig:FASER_eventformat}. 

The full event header is assigned to the event by the event builder. It contains information that identifies the event; it is composed by 13 words of variable size, which sum up to 352 bits (11 4-byte words), dominated  by the words corresponding to the data size, the event counter, an event identifier number and the time stamp. 
The readout-receiver event header is assigned by the DAQ receiver application. It contains similar information to the full event header, but relevant to the corresponding sub-detector. Each readout-receiver event header has a size of 288 bits (9 4-byte words).  The readout event header is assigned by the detector readout board directly, before the data are sent out. The content is specific to the sub-detector electronics.

Table~\ref{tab:data_size} displays the typical FASER data size per component. The total size is around 22~kBytes per event, dominated by the PMT readout (19~kBytes per event) due to the long readout window foreseen at least during early data taking.

\begin{table}[h!]
\begin{center}
\caption{Individual data fragment sizes and total size in kBytes. The per TRB payload size stated is the estimated average in the TI12 system during commissioning. The digitizer payload size is for all 16 channel readouts enabled and maximum readout window (1.2~$\mu$s as used during initial data taking).\\}
 \begin{tabular}{|c|c|} 
 \hline
 Component & Size [kBytes]  \\ [0.5ex] 
 \hline\hline
  TLB &  0.024 \\
  \hline
  TRB &  9 x 0.2 (1.8) \\
  \hline
  Digitizer &  19.2 \\
  \hline
  \hline
  Total payload&  21.0 \\
  \hline
  Total event size& 21.5 \\
  \hline
\end{tabular}
\label{tab:data_size}
\end{center}
\end{table}


\subsection{Read-out software: digitizer}\label{sec:daq_digi}
The digitizer system is controlled and read out in software by a custom C++ library built on top of the C++ driver library supplied with the SIS3153 card for communication over Ethernet. It provides an API for configuring all aspects of the SIS3153 and VX1730 boards based on a JSON configuration file. All configuration parameters are read back and checked after being sent to the boards. The data readout is accomplished by polling the FIFO buffer size until non-zero and then requesting the SIS3153 board to read and send the data of one event at a time. Multiple UDP packets are needed to transmit a single event and the SIS3153 board is configured to transmit all of the UDP packets corresponding to one event without waiting for an acknowledge signal from the server. Each UDP packet is numbered, so that the software can handle out-of-order packets as well as detect if a packet is lost. In case of a packet being lost, it is not possible to re-transmit the lost one as the data are not buffered after transmission, and the particular data fragment will be marked as corrupted. Reading just one event at a time ensures that the effect of such data loss is minimized. 

The library can be used both in a standalone application, allowing commissioning of just the digitizer board, and by the digitizer receiver module inside the full FASER DAQ system. In the latter case, the receiver module implements the state transitions needed by the DAQ, receives the configuration from the central run control system (see, next section) instead of a local JSON file and the received data fragments are transmitted to the event builder after adding the common DAQ fragment header.

\subsection{Read-out software: GPIOdrivers }
\label{sec:daq_gpiodrivers}


The communication with the GPIO board, which is used as the basis for the TRB and TLB, is encapsulated in 
a C++ library called \textit{GPIOdrivers}~\cite{gpiodrivers-sw}. The library provides a high level API tailored to the 
needs of the TRB and TLB. The serial communication with the hardware via USB and Ethernet links
using the GPIO specific protocol is completely encapsulated in this library. 
The architecture and main features of the API will be discussed in the following. 

\begin{figure}[h]
    \centering
    \includegraphics[width=0.7\textwidth]{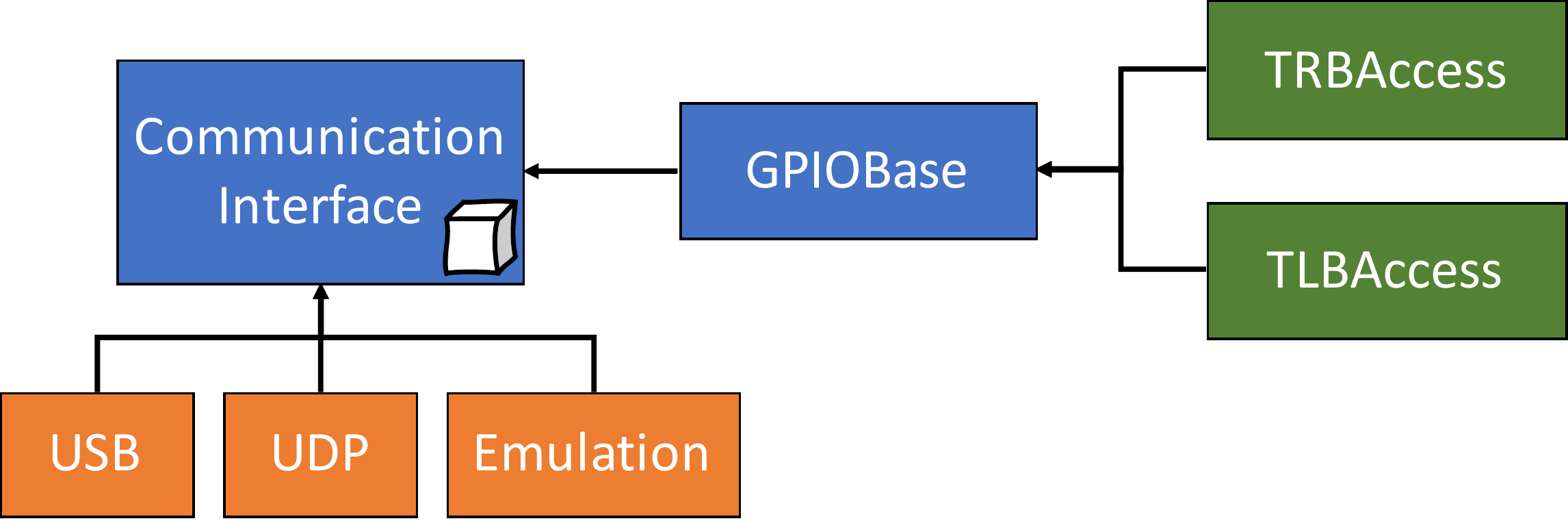}
    \caption{Overview of the GPIODrivers class architecture. The CommunicationInterface class is abstract. The classes that handle data transfer (UPD, USB and the emulation of those) derive from it. An instance of that class is also included in the GPIOBase class, which provides the low-level functionality, from where the two implementations (TLB and TRB) derive.  }
    \label{fig:GPIOdriversClassStructure}
\end{figure}

The class structure of the GPIOdrivers library is shown in figure \ref{fig:GPIOdriversClassStructure}.
It consists of three main parts: the encapsulation of the transport layer specific needs, the low level access to the GPIO board, and the TRB / TLB specific implementation of high level features. 
\begin{enumerate}
    \item On the transport layer the communication with the GPIO board is performed by sending and receiving 
    32-bit words. Two communication channels are needed by the GPIO board: one bi-directional for 
    configuring the board and querying status information and an additional unidirectional channel 
    from the GPIO board to the host PC, used to receive readout data.
    The corresponding functionality is defined in the abstract class \textit{CommunicationInterface}
    and implemented in three derived classes, each of them handling data transfer via the UDP network protocol, via USB or emulating the hardware. 
    
    \item The low level functionality of the GPIO board is accessible via the \textit{GPIOBase} class
    which holds an instance of the \textit{CommunicationInterface}. It encapsulates the communication 
    protocol of the GPIO board, allowing to send individual commands with arbitrary payload to the 
    GPIO board. The integrity of the transmitted data is confirmed by a checksum shipped with every command. 
    The hardware addresses of all available commands as well as all possible error codes
    are defined in this class for user friendly access. 
   Retrieval of the status and firmware version of the 
    GPIO board is also implemented here. 
    
    \item The high level interaction with the TRB / TLB firmware is implemented in two classes 
    derived from the \textit{GPIOBase} class: \textit{TRBAccess} and \textit{TLBAccess}.
    Both classes implement the access to the firmware functionality and registers provided by the 
    respective firmware implementations. Helper functions to 
    configure the GPIOBoards as well as default configurations for the TRB / TLB 
    are provided. In addition, the TRBAccess class also encapsulates the communication with the 
    attached tracker modules. 
    The GPIO driver software waits to receive readout data from the GPIO boards, which is then buffered, analysed for errors, and packed into 
FASER data fragments, if needed in standalone running. 
To maximize the data throughput the receiving and processing of readout data is performed in separate threads.
\end{enumerate}

The GPIODrivers package includes standalone programs to test the communication with the TRB / TLB boards, 
configure them and read out data. These programs can be use as templates for future projects.  
Beside standalone running of individual boards, the library is further employed in the tracker and trigger receiver modules within the context of the FASER DAQ framework (see section \ref{sec:daq_sw}). For the latter, as in the case of the digitizer, configuration settings are loaded from the central run control system and receivers issue control commands to the boards depending on the run control state (see section \ref{sec:control_config}). The received data fragments are transmitted to the event builder after adding the common DAQ fragment header.

\section{Controls and monitoring software}\label{sec:controls_monitoring}

Given the fact that the FASER experiment will be running on an operational model without shifters but with remote controls and monitoring of the detectors and TDAQ, special emphasis is given to the controls and monitoring infrastructure of the experiment. The run control and configuration software is summarised in section~\ref{sec:control_config}. The monitoring infrastructure put in place is  outlined in section~\ref{sec:monitoring} and finally, the DCS software infrastructure, which uses that of the large LHC experiments as a guideline, is presented in section~\ref{sec:dcs_sw}.

\subsection{Run control and configuration}\label{sec:control_config}
The low-level process control is implemented in DAQling on top of the `supervisord' library \cite{supervisord}. Each process is started / stopped using supervisord and follows a common finite state machine (FSM) to control and synchronize the configuration and data taking steps. The control is steered by a Python library provided by DAQling, which communicates with the supervisord library and the individual processes using XML-RPC. This library is also responsible for transmitting the individual configuration for each module in the form of a JSON object, again sent using XML-RPC.

\begin{figure}[ht]
    \centering
     \includegraphics[width=0.85\textwidth]{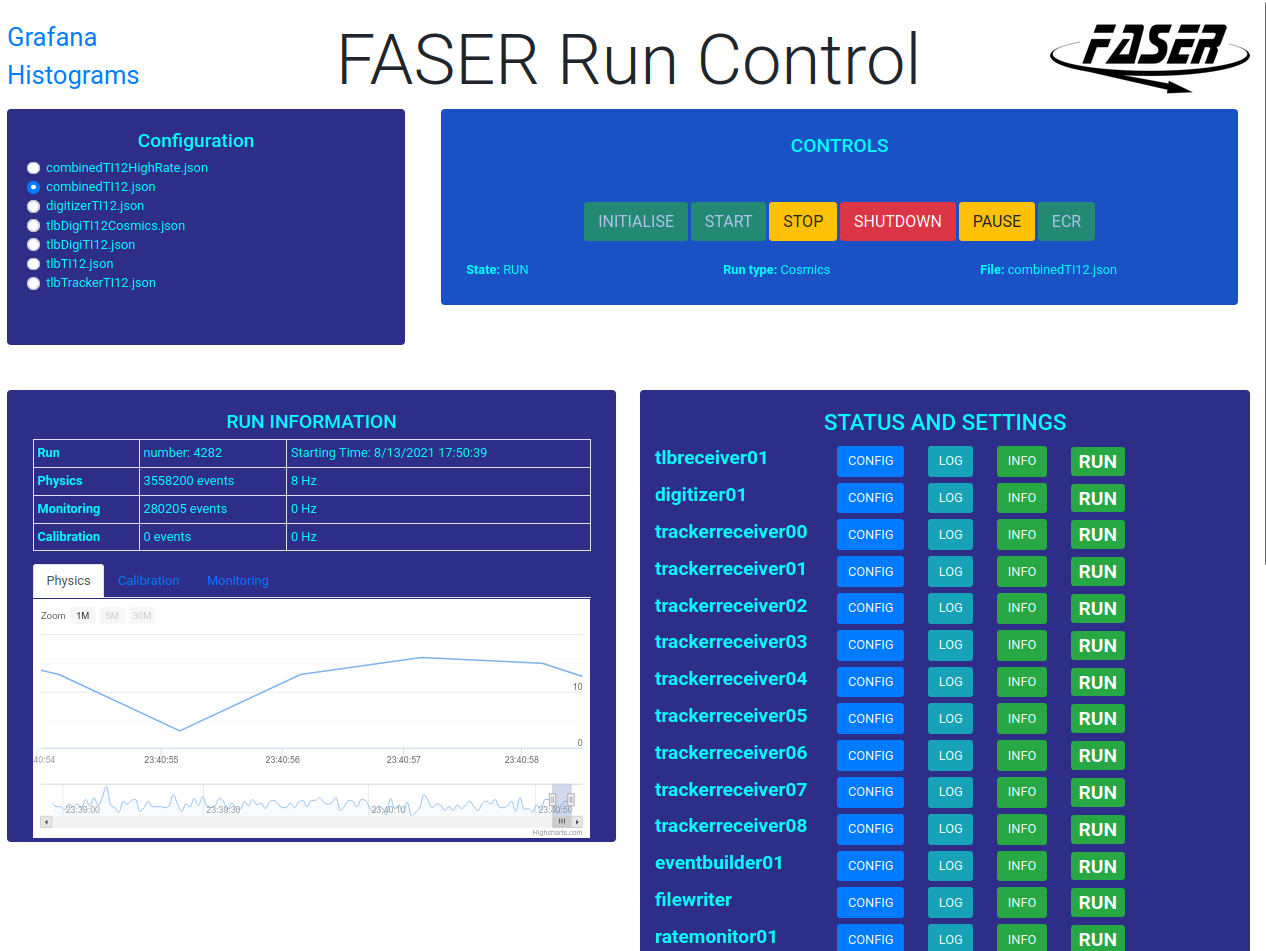}   
    \caption{Example screen shot of the RCGui web application during cosmic ray data taking with the FASER detector.} 
    \label{fig:RCGui}
\end{figure}

On top of the low-level control library, a FASER-specific Run Control web application (RCGui) has been built using the Python-based Flask web framework to allow easy control of the DAQ from remote machines as illustrated in figure~\ref{fig:RCGui}. The web application allows the user to select between different pre-defined run configurations as well as editing all settings directly in the web browser. The application provides control to cycle through the FSM states and monitor the progression of each process. It also provides easy access to the log file of each application as well as the monitoring data, as described in the next section. At the start of each run, the run control assigns a new sequential ``Run number'' to the data that will be taken and archives the full configuration used into an Oracle database along with a user supplied comment.
Similarly when a data taking run is stopped, it allows for a user comment to be added to the run record in Oracle together with the total number of events recorded.

\subsection{Monitoring}

Extensive monitoring has been put in place to monitor data flow and data variables at every step of the data acquisition.
Two types of monitoring exist: (i) monitoring of a single variable, or metric, over time for monitoring of conditions; (ii) monitoring of histogrammed distributions of variables for data quality control.

\paragraph{Monitoring of metrics.}
Metrics are defined and regularly published in every DAQ process; for example, the total number of physics events that have been read out is a common metric for all DAQ processes.
Metrics and  handling of those are part of the \DAQling{} infrastructure. Different metric types determine the final arithmetic applied upon publication, such as the computation of the average, the rate or current state of a variable over the publishing time interval.
Metrics are published at regular intervals to \InfluxDB{}~\cite{influxDB} and \Redis{}~\cite{redis}, and viewed in \Grafana{}~\cite{grafana} dashboards and the Run Control interface.

\begin{figure}[h]
    \centering
    \includegraphics[width=\textwidth]{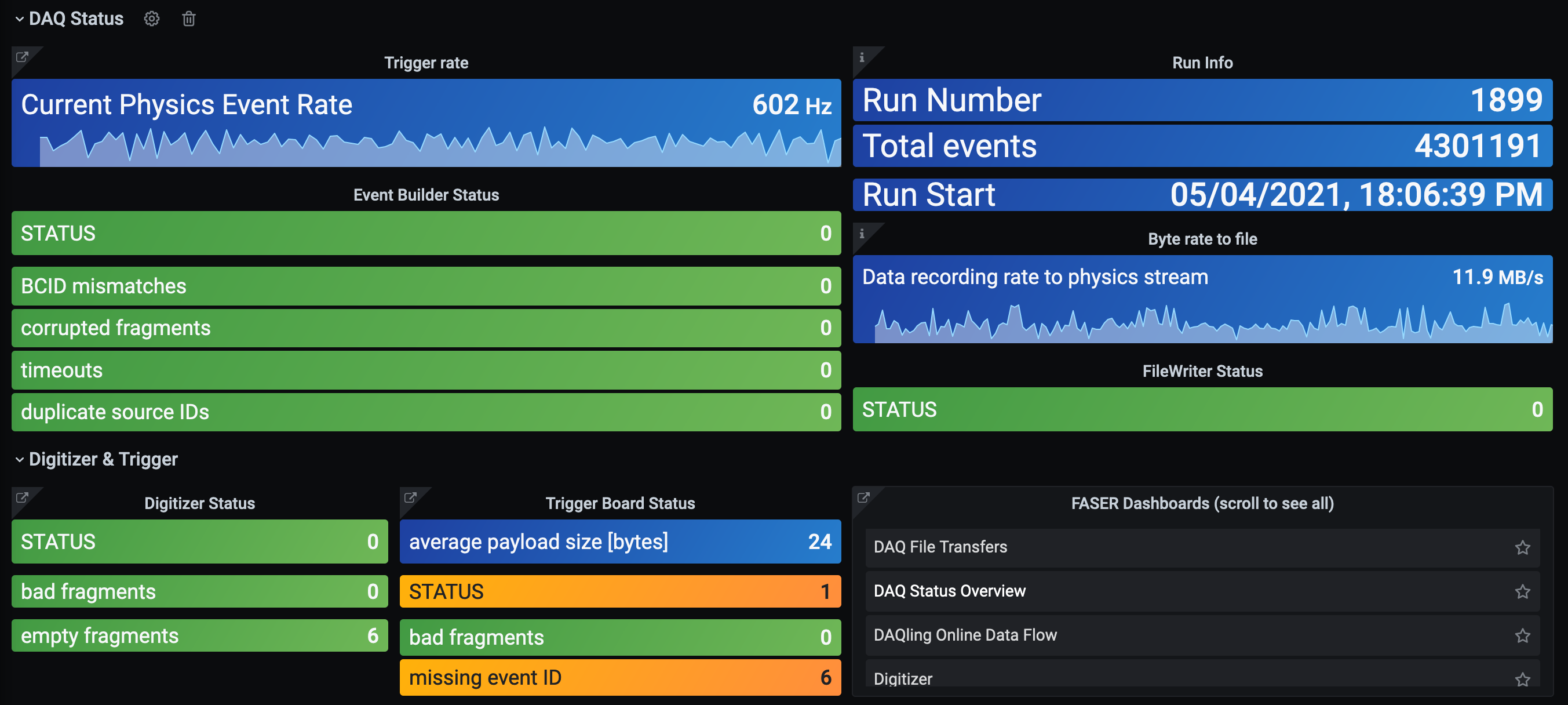}
    \caption{\Grafana{} dashboard of the DAQ status overview for an ongoing data taking run. Metrics of the trigger and data recording rate, DAQ module status and total error counts are displayed. In addition, current run details are shown.}
    \label{fig:graf_DAQStatus}
\end{figure}

\begin{figure}[h]
    \centering
    \includegraphics[width=\textwidth]{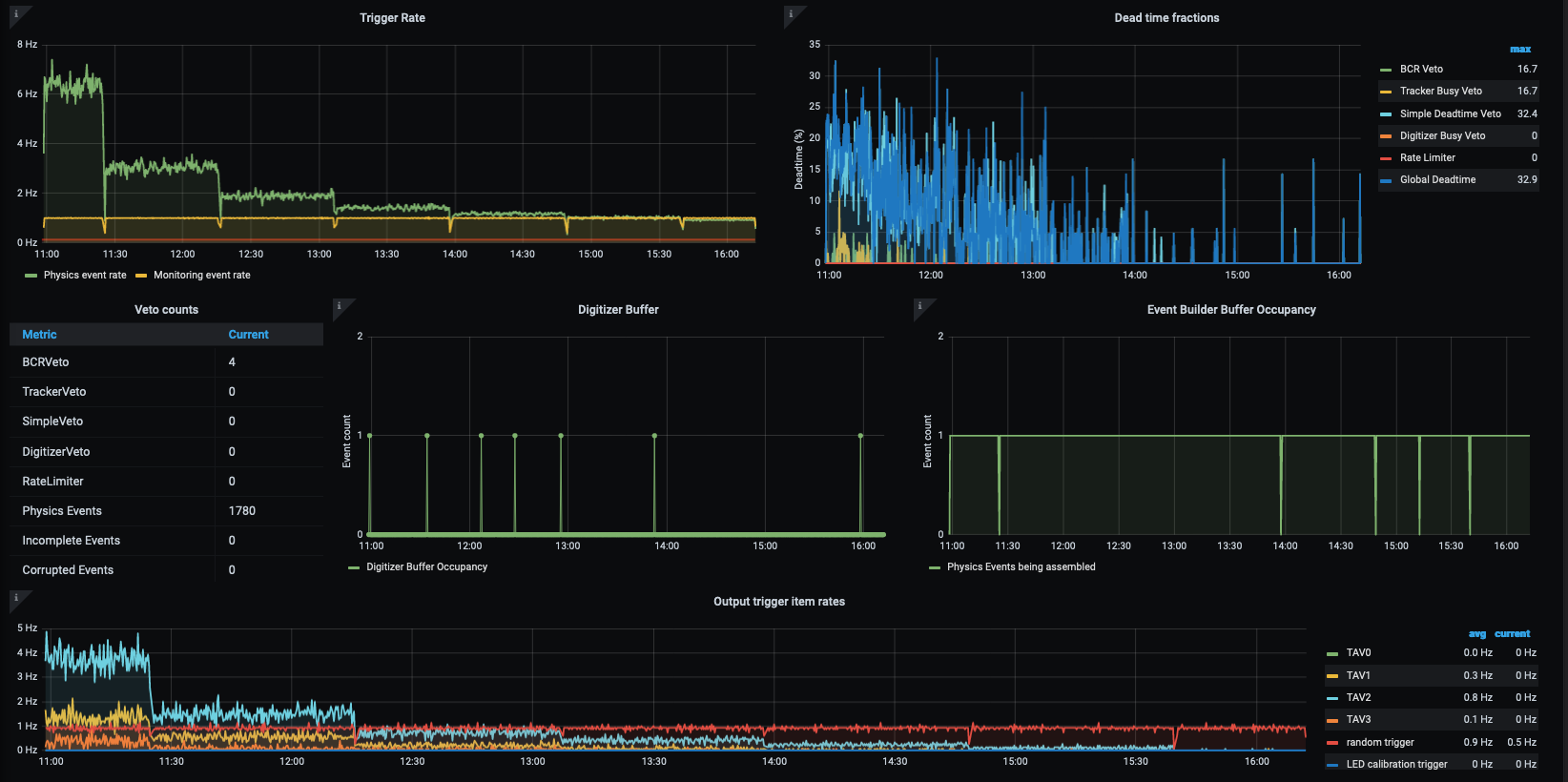}
    \caption{\Grafana{} dashboard of rate and buffer information during PMT high voltage scans: the total physics and monitoring event rate, deadtime fractions, buffer occupancies in the digitizer and software event builder, and rates of individual trigger items are monitored.}
    \label{fig:graf_FASERTrigger}
\end{figure}

Examples of \Grafana{} dashboards are shown in figures~\ref{fig:graf_DAQStatus} and \ref{fig:graf_FASERTrigger}; they demonstrate, respectively, monitoring of the overall current status of the data acquisition system, and the time trends in rates and buffer occupancy.

\begin{figure}[th]
    \centering
    \includegraphics[width=\linewidth]{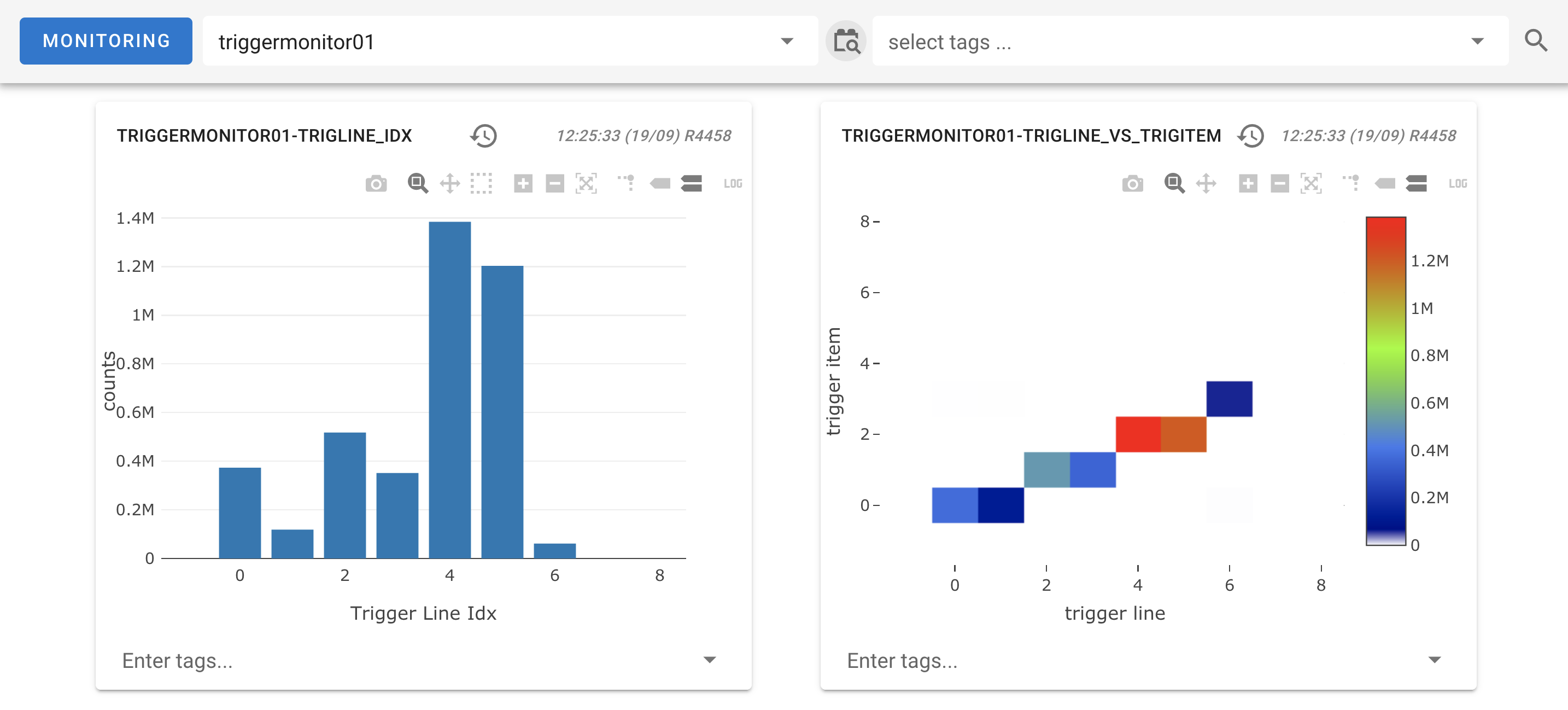}
    \caption{Histogram monitoring home page: the monitored histograms are displayed in a grid and they have their own options (ID, historic page icon, timestamp and custom tags). An example of a one-dimensional histogram is shown on the left and of a two-dimensional histogram is shown on the right.}
    \label{fig:monitoring_home_page}
\end{figure}

\paragraph{Monitoring of histograms.}
Histograms are useful for monitoring extreme or rare outliers.
Dedicated DAQ monitor modules receive copied event data from the event builder. The monitors  unpack events to visualise event features in detail via histograms and metrics.
Histograms are regularly published to the \Redis{} database, from which a separate histogram polling application extracts the histograms to display on a web interface.

The web interface is written in Javascript with the Vue~\cite{vue} framework and communicates with a Flask~\cite{flask} server written in Python. A histogram published in the database can then be retrieved by the Flask server and sent to the web application. From there, the Javascript library Plotly.js  displays the different types of histograms in the browser. Histogram tagging and timestamps are used for easy navigation through the interface. Monitoring data arrives promptly to the histogramming tools and a histogram update frequency down to 5~s has been tested. The update frequency to be used for the data taking with beams will be fixed with operational experience. Example monitoring histograms are shown in figure~\ref{fig:monitoring_home_page}.
\label{sec:monitoring}

\subsection{DCS}

\begin{figure}[h]
    \centering
    \includegraphics[trim=1cm 0.5cm 1.5cm 3.5cm, clip=true, width=0.8\textwidth]{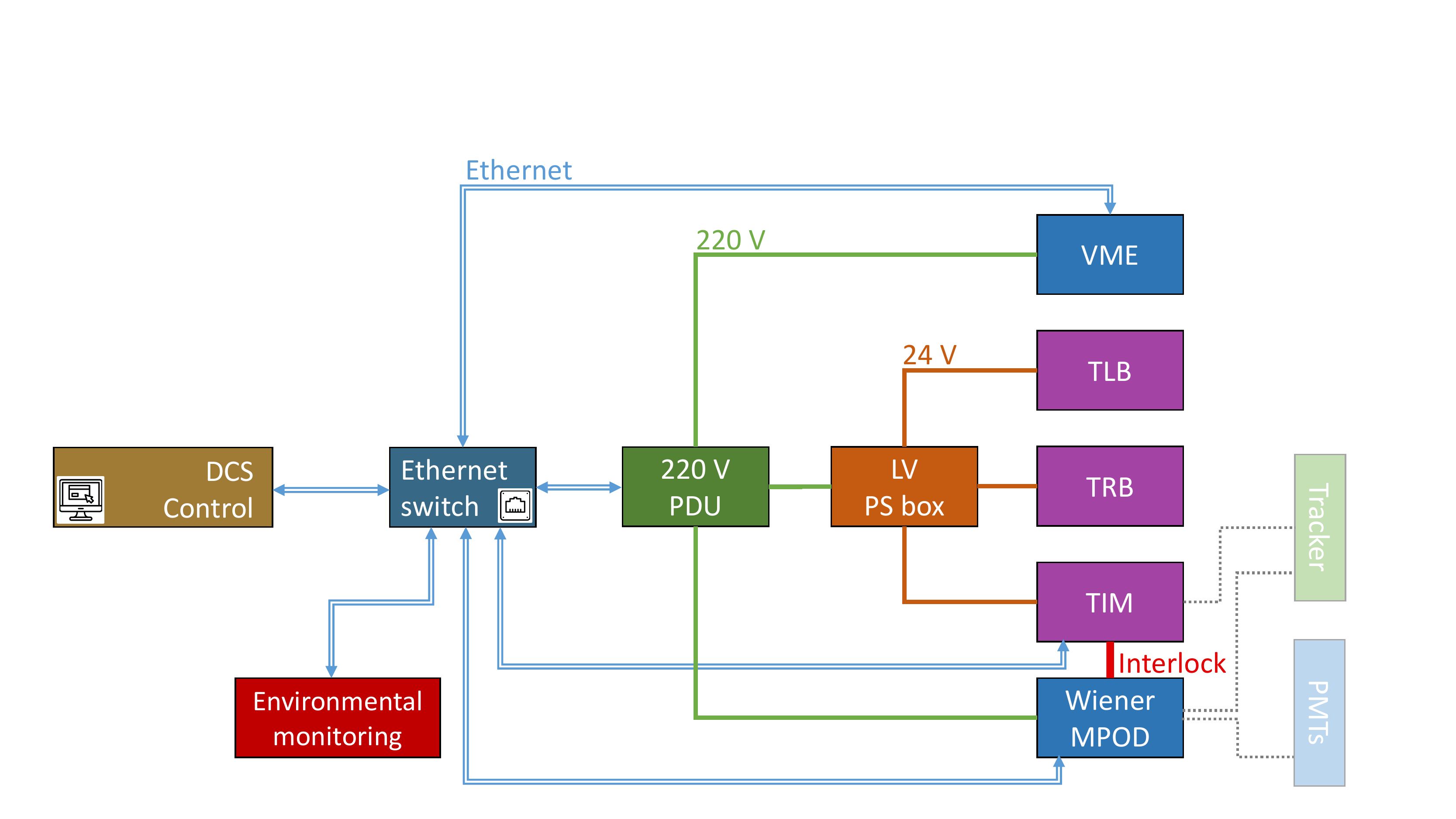}
    \caption{The DCS components of the FASER experiment and the connections between them. The DCS connections to the detector are also shown. 
    }
    \label{fig:DCS}
\end{figure}

The FASER DCS is designed to ensure  the safety of the detector during the different operation modes, to monitor the operating parameters of the detector’s power system and its environment, and to control and configure the detector's operational states.
The software back-end of the  DCS is implemented with the SIEMENS Simatic WinCC Open Architecture, which is a SCADA system used at CERN by the LHC experiments, as already introduced in section~\ref{sec:TDAQ_global}. The DCS components of the experiment and how they connect with each other are shown in figure~\ref{fig:DCS}.

The communications between the DCS back-end and the devices is routed through a commodity switch using standard 1~Gbit/s Ethernet, identical to the DAQ switches and network described in section~\ref{sec:daq_hw}. The Ethernet network of the devices is secured from the CERN general network. 

The items controlled and monitored by the DCS can be grouped in the following categories:
\begin{enumerate}
    \item Environmental parameters: The temperature and humidity of the tracker is monitored through a custom electronic board (TIM board). Data is transmitted to the DCS back-end via the industry standard Modbus protocol~\cite{modbus}.
    \item Tracker safety interlock: The TIM board also serves as a hardware-based safety interlock for the tracker. It is configured and monitored via Modbus by the DCS back-end.
    \item HV and LV system: The main HV and LV system is based on the Wiener MPOD system~~\cite{wiener_mpod} and is monitored and configured by the DCS via the OPC-UA.
    \item Power Distribution Units: A commercially available Power Distribution Unit (NetIO PowerPDU 4C~\cite{PDU}) is controlled by the DCS via Modbus. 
\end{enumerate}

To ensure the safety of the detector, the DCS triggers alerts and automatic emergency procedures. The DCS also implements autonomous integrity checks and diagnostics and provides a user interface to inform FASER on-call experts on the state and status of the experiment. 

\begin{figure}[h]
    \centering
    \includegraphics[width=\textwidth]{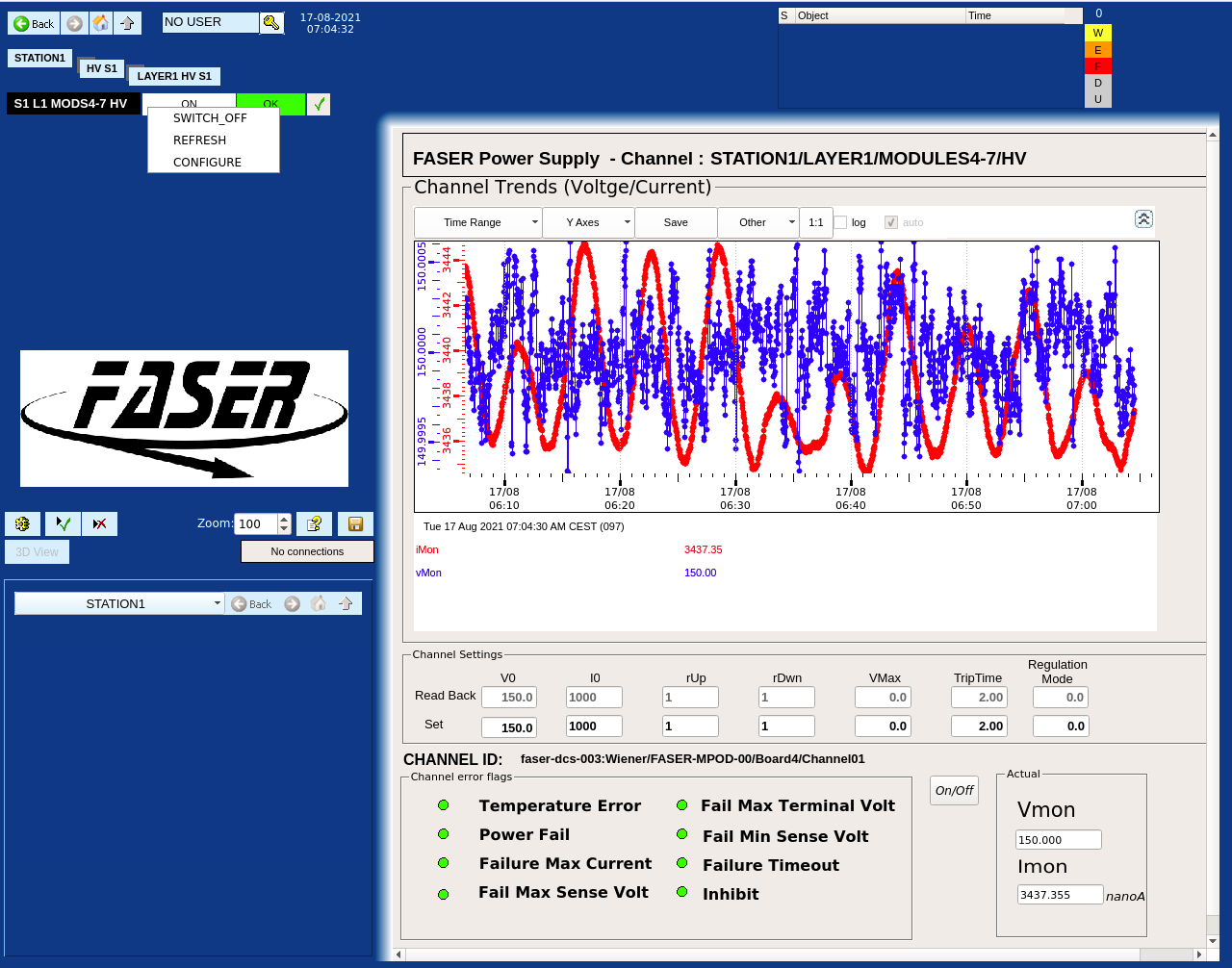}
    \caption{A DCS FSM panel via which the low and high voltage of a tracker station is controlled.}
    \label{fig:dcs_fsm}
\end{figure}

Transitions between the different detector operational states are implemented as a FSM (see figure~\ref{fig:dcs_fsm}). Transitions are initiated either by a user input or by an autonomous protocol. The action of the transition is initiated by propagating a transition signal through a sub-system hierarchy which in turn triggers a series of autonomous protocols that configure the necessary parameters to achieve the transition between the different operational states. 

The DCS also provides a user interface for configuring operational parameters of the power and interlock systems and an interface to a database to which operation-parameter configurations can be saved and extracted from. The DCS monitors approximately 600 parameters at a 1~Hz rate, collecting data at a rate of 1.3~kBytes/s. 

\begin{figure}[h]
    \centering
    \includegraphics[width=\textwidth]{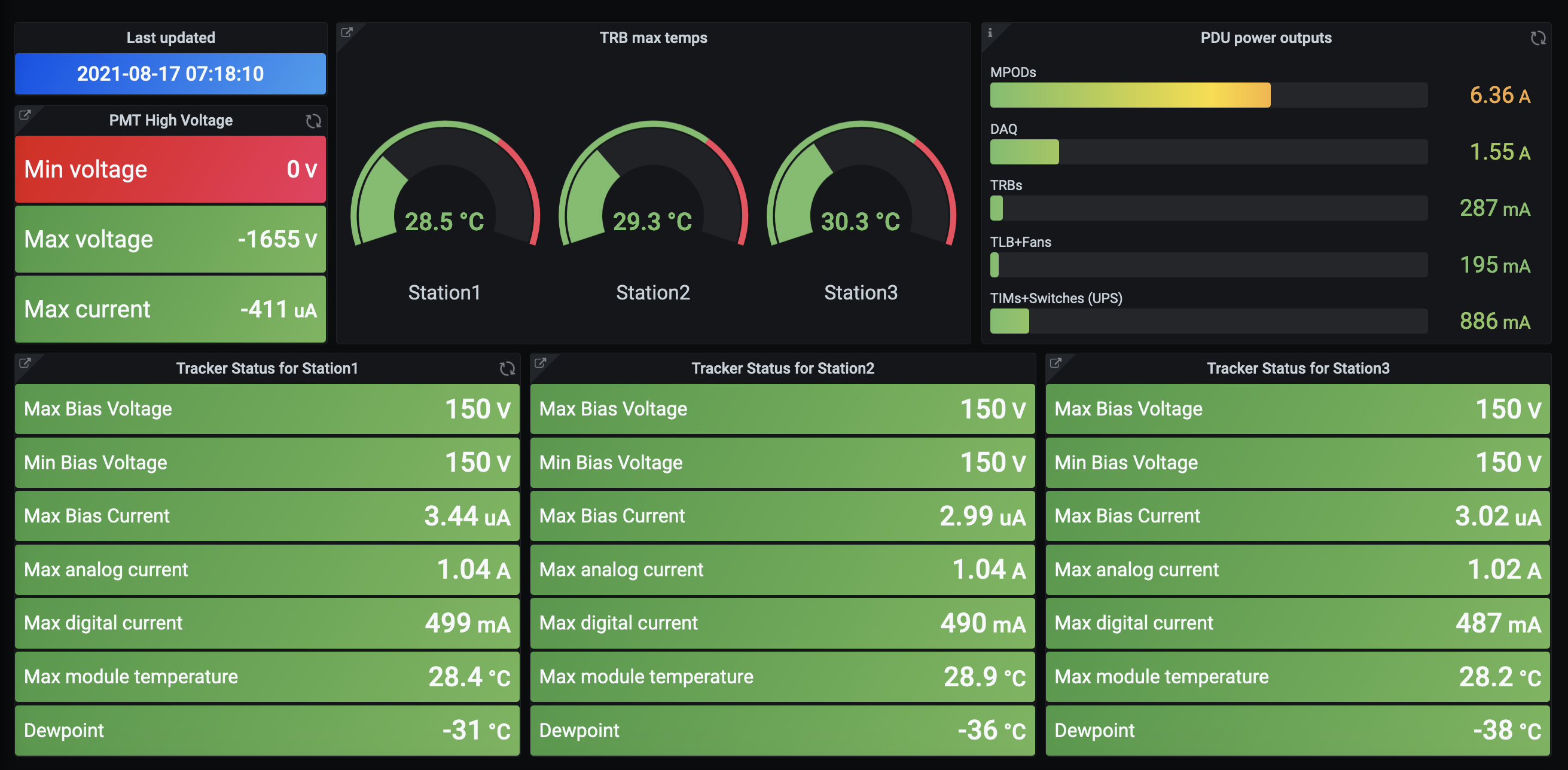}
    \caption{\Grafana{} dashboard of the DCS status overview during an ongoing combined system run. Metrics such as current tracker and PMT voltages, temperatures and power output are shown.}
    \label{fig:graf_dsc_status}
\end{figure}

The DCS also records time-series of the detector parameters, as well as the operational state and the detector status. The time series are archived for later retrieval in an Oracle database, which ensures the quality of the data collected at a particular time. Figure~\ref{fig:graf_dsc_status} shows an example of DCS values retrieved and monitored in \Grafana{}.

\label{sec:dcs_sw}

\section{Commissioning and system performance}
\label{sec:standalone_commissioning}

This section of the paper describes the TDAQ system commissioning in the three FASER commissioning phases introduced in section~\ref{sec:intro}. The initial step of the system commissioning has been the qualification of the hardware and software components, which was achieved with basic testing and confirmation of the functionality. The subsequent step has been the assessment of the system performance, both of each component individually and of the full system combined. Results from both these steps are reported in the following sections, first for each component separately in sections~\ref{sec:digitizer_com}--\ref{sec:TRB_com}, for the full DAQ system in section~\ref{sec:DAQ_com} and then for the full combined system with the complete detector connected in section~\ref{sec:combined_com}.
For all tests described in this section the system was sized up to what is required for the FASER experiment without yet accounting for FASER$\nu$, i.e. for three tracking stations and three scintillator stations.

\subsection{Digitizer standalone commissioning}\label{sec:digitizer_com}
The CAEN digitizer card and the Struck VME-to-Ethernet card were tested together using both signal generators and real PMT pulses with the digitizer self-triggering and the data read out over the Struck card either connecting directly to the Ethernet port on a Linux-based server or over a local network switch. 

\subsubsection{Digitizer trigger timing}

A critical aspect of the digitizer trigger functionality is the trigger latency and in particular the jitter in the output trigger signals as well as their synchronicity. This was tested with different types of signals, with the most accurate test being a square pulse sent to multiple inputs with small (1--4~ns) relative delays to ensure a fixed expected ordering of the trigger signals. The trigger output signals are routed back to other input channels on the digitizer and the trigger signal arrival times can be compared to the arrival time of the pulses. During this testing, one board was found to have an additional 8~ns jitter on some channels in an irreproducible fashion. Once discovered, the board was promptly exchanged by the manufacturer.

\subsubsection{Readout performance}

The maximum readout performance of different configurations for the readout are tested by filling the digitizer buffer with 1000 triggered events and measuring the time needed to read those out. The best performance is found when using the highest speed VME64x readout protocol (2eSST320) and the VME-to-Ethernet card configured to send up to its maximum 32 Ethernet packets (with up to 1440 bytes of VME data per packet) for each data request sent by the readout application. As UDP is used for the network data transfer, data loss will occur in case a packet is dropped on the network as the VME-to-Ethernet card does not have the capability to resend a lost packet when more than one packet is sent per readout request. The data requests are therefore always done for a single full event or a small multiple of events, so that only that or those events are affected by any data loss. The amount of data requested should also stay below the maximum possible data transfer for a single data request ($32\times1440$~bytes$=45$~kBytes) in order to have the maximum resilience in handling packet drops.

\begin{figure}[h]
  \centering
  \includegraphics[width=0.8\textwidth]{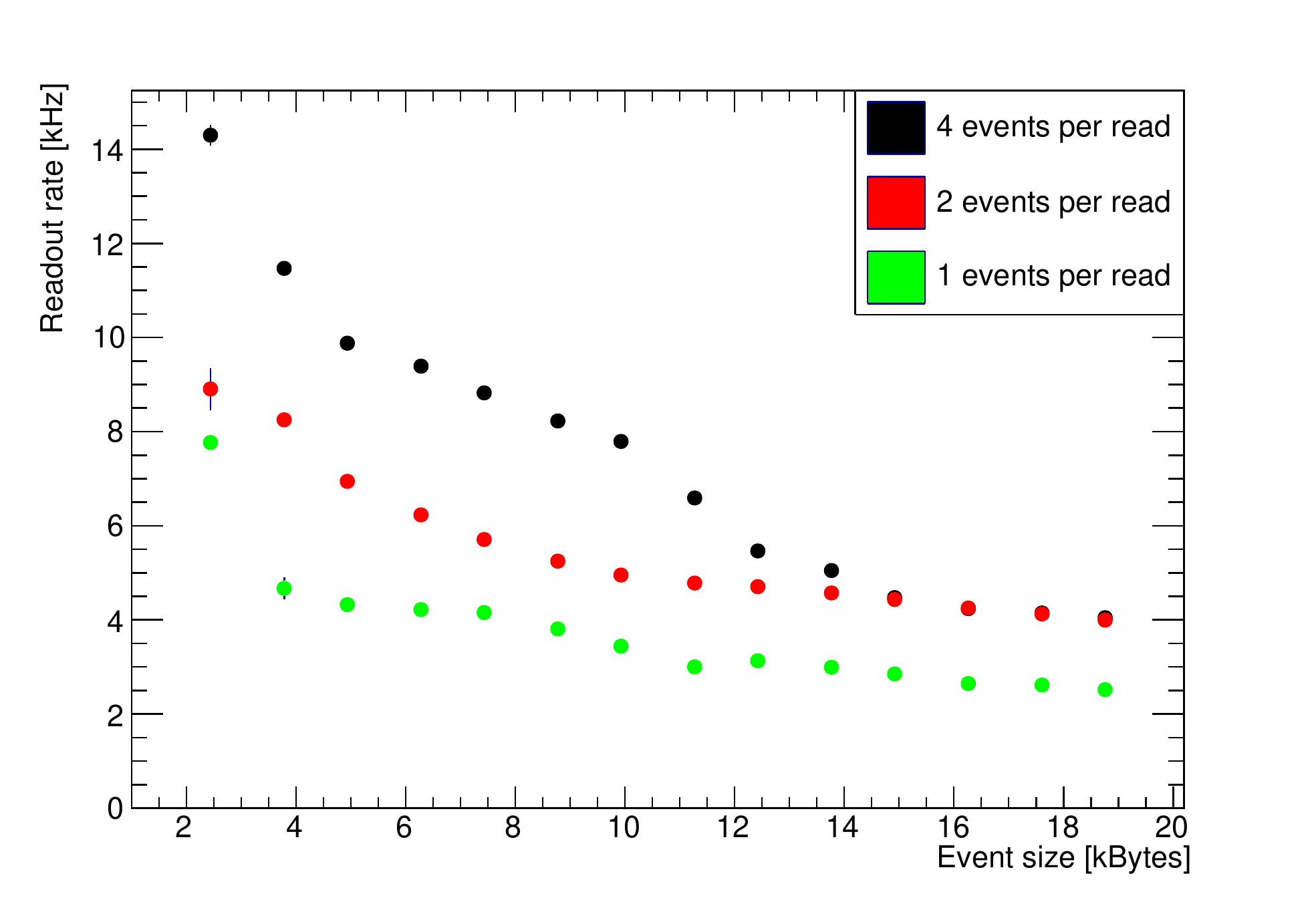}
  \caption{Maximum achievable read out rate for the digitizer as a function of the event size for three different groupings of the events.}
  \label{fig:digitizerRate}
\end{figure}

A maximum readout rate at the nominal 19~kBytes fragment size of more than 2.5 kHz has been demonstrated when reading out a single complete event per request as shown in figure~\ref{fig:digitizerRate}. This corresponds to a maximum bandwidth usage of almost 50~MBytes/s or about half the maximum network link speed. 
Higher readout rates are possible if the readout window and thus event size is reduced, but the effective bandwidth goes down due to the additional overhead of pulling individual events. The readout rate can be further improved by requesting multiple events per readout request as long as the total data size fits inside the maximum 45~kBytes per readout request. With that a maximum readout rate of almost 4 kHz can be achieved for the nominal digitizer fragment size. A further increase can be achieved by switching to using the larger Jumbo Ethernet packets (a data payload of up to 7168 bytes per packet instead of 1440 bytes) with almost 5~kHz reached in standalone testing when requesting four events per readout request.

In combined running, the full readout rate has not been achieved as the available bandwidth is reduced by the additional TRB and TLB readout data. Furthermore the current digitizer receiver application is single-threaded meaning that a single thread is both receiving the data, decoding it and preparing it for transmission to the event builder, adding latency between receiving one event and requesting the next one. Running with  the digitizer in the production DAQ system, the maximum readout rate when requesting just one event per request is measured to be 2.2~kHz for the nominal event size. This increases to about 2.8~kHz when requesting two events per readout request. It was not possible to test with Jumbo frames in the production system due to the lack of a direct network connection, but in the test system the rate increased to 3~kHz when requesting four events per readout request. The multiple event requests slightly increase the risk of data loss and are not expected to be needed at the nominal FASER trigger rates, but could be deployed in case the physics trigger rates are higher than expected or for special high rate data taking.



\subsection{TLB standalone commissioning}
\label{sec:TLB_com}
The functionality of the trigger logic board described in section~\ref{sec:hw_tlb} is tested in a lab setup 
in which a signal generator is used to supply various configurations of signals to the CAEN digitizer and the digitizer trigger outputs are fed to the TLB.
The results are verified by studying the signal inputs and outputs on an oscilloscope and / or analysing the data output from the board.


\subsubsection{Signal timing}

\begin{figure}[t]
  \centering
  \includegraphics[width=0.7\textwidth]{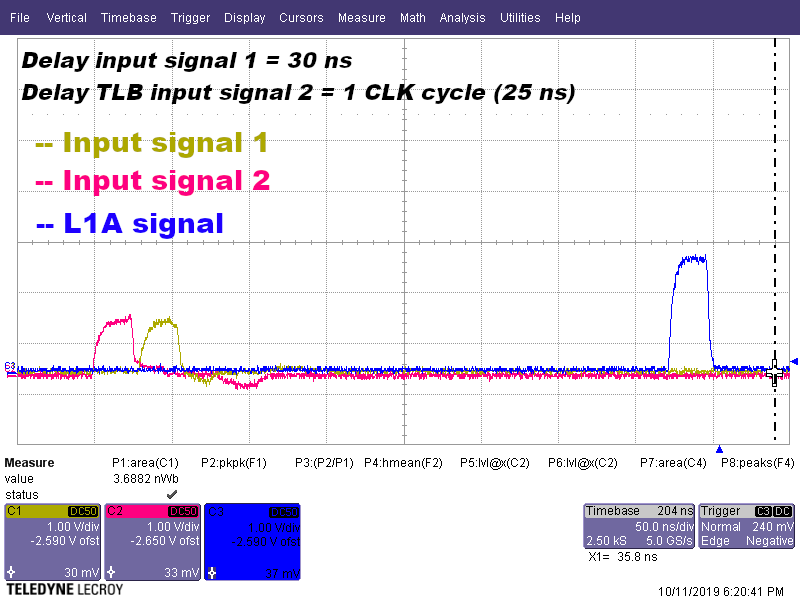}
  \caption{An oscilloscope image of two trigger input signals sent by the digitizer. The TLB is set to only trigger on a coincidence of the two input signals. Input signal 1 is purposefully delayed by 30ns (over 1 clock cycle). A delay of 1 clock cycle is applied on input signal 2 on the TLB, which has resynchronised the input signals to give rise to an L1A.}
  \label{fig:scope_inputdelay_test}
\end{figure}

Input signals to the TLB can be individually timed via two functionalities to ultimately synchronise all input signals stably: a delay in increments of 1 BC up to 3 BCs total, and the option to sample an input signal on the falling or rising edge of the LHC clock.
The input delay is tested by splitting the signal from a signal generator and purposefully delaying one of the signals by 30~ns (just over one LHC clock cycle) before feeding them to the digitizer. The TLB trigger \LUT{} is configured to only trigger on the coincidence of the two input signals from the digitizer. Figure~\ref{fig:scope_inputdelay_test} shows two input signals, with a difference in arrival time of 30 ns, and the L1A signal sent out by the TLB. The L1A signal appears only once an input delay of 1 \BC{} is applied to the early signal, thereby showing that the TLB can successfully synchronise trigger input signals that would arrive at different times due to differences in signal processing, cable lengths and time-of-flight. The test is repeated for time differences of up to 3 clock cycles.

\subsubsection{Rate control}

The rate on the TLB is controlled either by prescaling individual trigger items or via the vetoing of triggers due to several sources described in section \ref{sec:hw_tlb}.

\begin{figure}[t]%
  \begin{subfigure}[~Prescale=1]{
    \centering
    \includegraphics[width=0.3\textwidth]{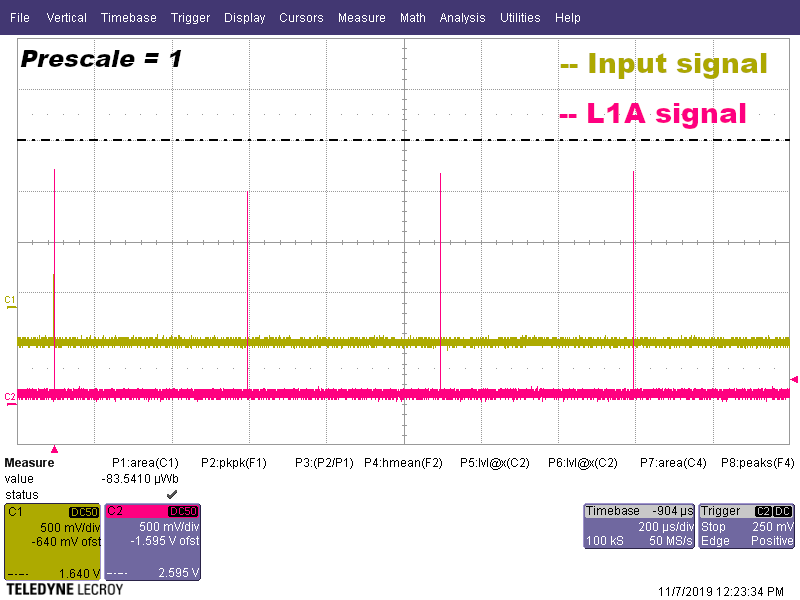}
    \label{fig:scope_tlb_prescale1}}
  \end{subfigure}%
  \begin{subfigure}[~Prescale=2]{
    \centering
    \includegraphics[width=0.3\textwidth]{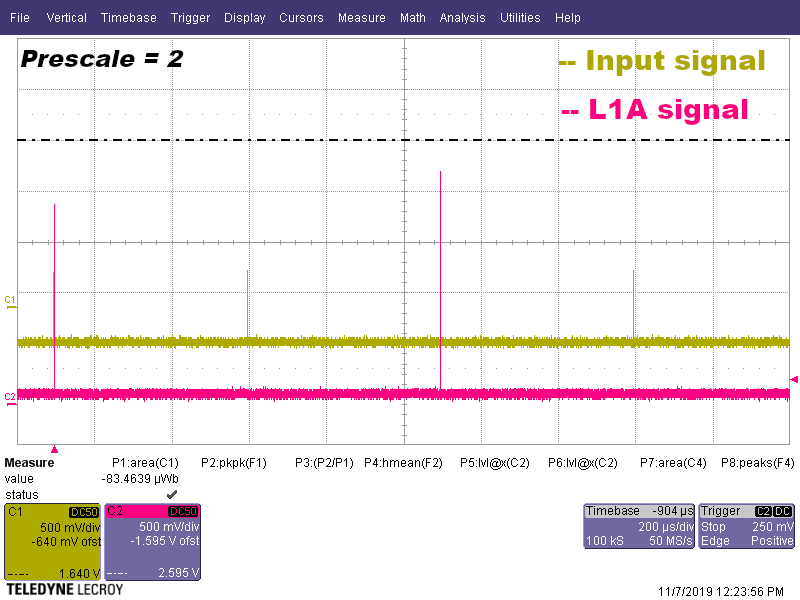}
    \label{fig:scope_tlb_prescale2}}
  \end{subfigure}%
  \begin{subfigure}[~Prescale=3]{
    \centering
    \includegraphics[width=0.3\textwidth]{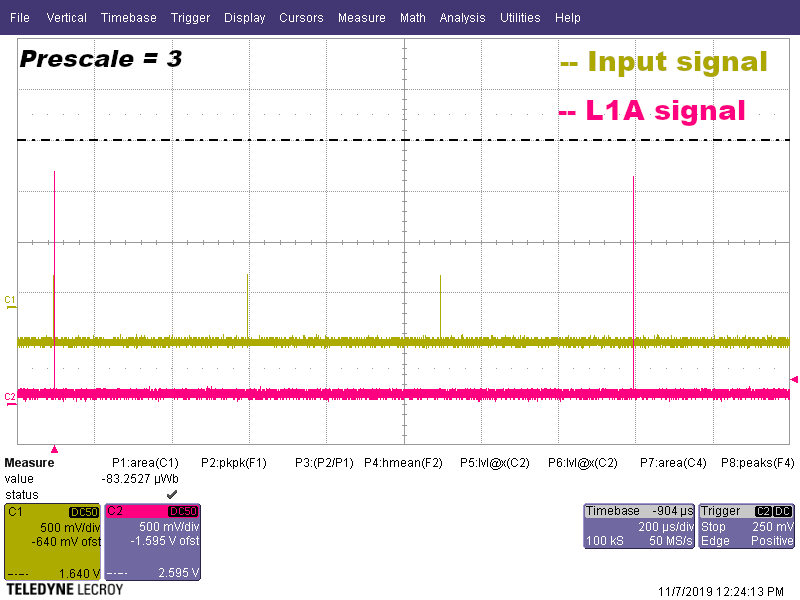}
    \label{fig:scope_tlb_prescale3}}
  \end{subfigure}%
  \caption{An oscilloscope image of a 2 kHz input signal sent from the digitizer to the TLB, and the subsequent TLB L1A signal, for different TLB prescale settings. The L1A signal appears only for every \textit{nth} input signal where \textit{n} is the prescale setting.}
  \label{fig:scope_prescale_test}
\end{figure}

The prescale functionality is tested on a single 2 kHz input signal from the digitizer to the TLB.
Oscilloscope images of the input signal and the TLB L1A signal are shown in figure \ref{fig:scope_prescale_test} for prescale settings of 1, 2 and 3. It is shown that only every \textit{nth} trigger gives rise to an L1A signal where \textit{n} is the prescale setting.

\begin{figure}[t]%
  \begin{subfigure}[~Simple deadtime=0 BC]{
    \centering
    \includegraphics[width=0.48\textwidth]{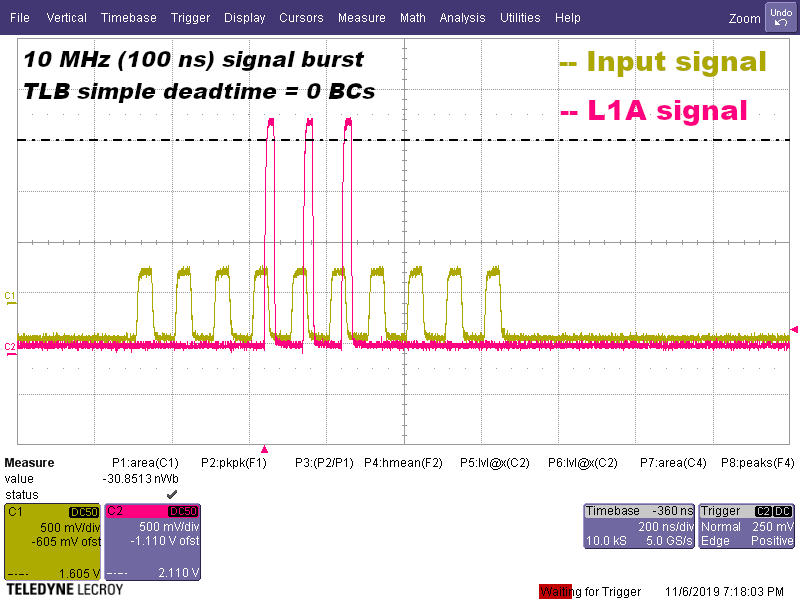}
    \label{fig:scope_tlb_deadtime0}}
  \end{subfigure}%
  \begin{subfigure}[~Simple deadtime=10 BCs]{
    \centering
    \includegraphics[width=0.48\textwidth]{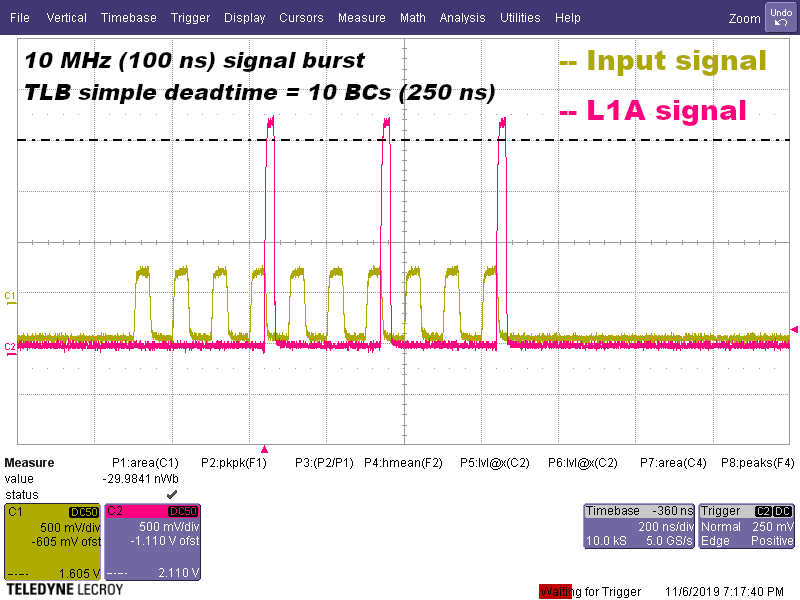}
    \label{fig:scope_tlb_deadtime10}}
  \end{subfigure}%
  \caption{An oscilloscope image of a 10 MHz 10-pulse signal burst sent from the digitizer to the TLB, and the subsequent TLB L1A signal, for different TLB simple deadtime settings. For a simple deadtime of 0 BC, the L1A appears for 3 consecutive signals. For a simple deadtime of 10 BCs (250~ns), the 2 input signals following an L1A are vetoed. Only 3 L1As appear in total for a pulse burst due to the TLB rate limiter.}
  \label{fig:scope_simple_deadtime_test}
\end{figure}

\begin{figure}[t]
  \centering
  \includegraphics[width=0.7\textwidth]{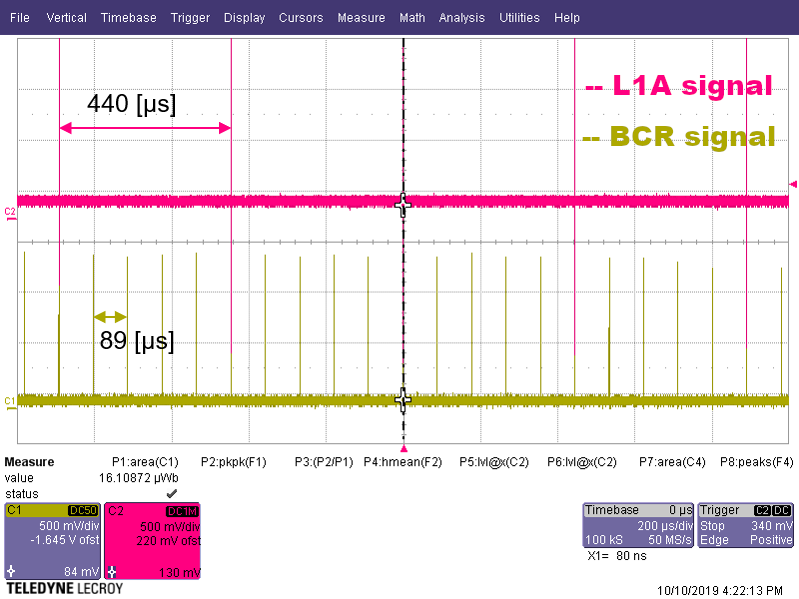}
  \caption{An oscilloscope image of the \BCR{} signal used as a trigger signal to the TLB, and the subsequent TLB L1A signal. Only every 5th BCR signal is shown to give rise to an L1A, as the TLB rate limiter has limited the triggers to 1 L1A every 5 orbits.}
  \label{fig:scope_ratelimiter_test}
\end{figure}

Three veto sources asserted by the TLB itself are tested in the lab setup, namely the simple deadtime veto, the rate limiter and the \BCR{} veto.
The simple deadtime veto is tested by supplying a 10~MHz 10-pulse signal burst to the digitizer-TLB system, as shown in figure~\ref{fig:scope_simple_deadtime_test} for a simple deadtime veto window of 0 and 10 \BCs{}. While the former shows three L1A signals every 100~ns, the latter shows three consecutive L1A signals for every third input signal 300~ns after. The 2 input signals in between are vetoed as they lie within the 250~ns shadow of the simple deadtime veto. The results further demonstrate the rate limiter functionality, which has not allowed more than 3 L1A signals per burst, or per orbit.
The rate limiter functionality is demonstrated again in figure~\ref{fig:scope_ratelimiter_test}. In this case the BCR signal from the TLB is fed into the digitizer. The rate limiter regulates the rate via a counter that increases upon an L1A and decreases by 1 on every 5th orbit. The veto sets in when the counter is above 3. Hence, the oscilloscope shot shows an L1A only for every 5th orbit signal, when the counter is decreased to 2 briefly, while the next 4 signals are vetoed.

\subsubsection{Further trigger features}

As described in section \ref{sec:hw_tlb}, the TLB has the capability to generate its own triggers, either at fixed or pseudo-random rate. 
Table~\ref{tab:randomtrig} lists the average random trigger rate over 1 million triggered events for 8 different available settings on the TLB. The random trigger is shown to have a triggering range of 15~Hz -- 1.2~kHz. 

\begin{table}[h!]
\begin{center}
\caption{Measured random trigger rate over 1 million events for all 8 random trigger settings available on the TLB.\\}
\begin{tabular}{|c|c|c|c|c|c|c|c|c|} 
  \hline
  \multicolumn{9}{|c|}{Random trigger average rate} \\
  \hline\hline
 Setting & 0 & 1 & 2 & 3 & 4 & 5 & 6 & 7 \\ 
 \hline
 Measured avg. rate [Hz] & 1.2k & 500 & 180 & 130 & 90 & 40 & 20 & 15 \\ 
 \hline\hline
\end{tabular}
\label{tab:randomtrig}
\end{center}
\end{table}

Other aspects are also verified; these include the BCR veto functionality and the expected BCID distribution in randomly triggered events, which is confirmed to be flat for BCIDs between 1 and 3555 and empty in the BCR veto window in BCIDs between 3556 and 3564. An essential functionality of the TLB is the coincidence logic, which leads to the final L1A; this is tested using multiple input signals and various LUT configurations.


\subsection{TRB standalone commissioning}
\label{sec:TRB_com}

A detailed description of the functionality of the TRB can be found in section~\ref{sec:hw_trb}. An overview of the readout software is given in section~\ref{sec:daq_gpiodrivers}. Detailed tests to qualify the hardware and the communication with it have taken place. These include tests of the serial communication with the TRB including access to all registers, of the communication with tracker modules and the readout of the module data. The performance of the readout has also been verified with a similar set-up and procedure as for the qualification of the hardware.


\subsubsection{Hardware communication}


The communication with every TRB is tested by first 
reading the firmware version from the hardware. The hardware addresses of the configuration registers depend 
on the firmware version. If this is successful all configuration registers are written with 
defined values that are compared to the ones returned upon reading from the registers. If no errors 
are identified the communication test is passed. 

Every TRB can connect to up to eight tracker modules. Every tracker module may use two individual clock and 
data signal paths. 
The assignment of channel numbers to hardware connectors is initially tested by observing the clock and 
data hardware signals on the TRB connectors. Similarly the assignment of both clock and data lines 
to every module is tested in comparison to the one selected in the TRB configuration.  
The basic communication with the attached tracker modules is tested by configuring the modules and
checking the current consumption of the tracking modules, which change after a successful configuration. 
After that, the main configuration register of every module can be read back and compared 
to the expected value. If both tests are passed without errors the communication with the modules is 
established. 

Lastly the data readout from the tracker modules is tested using calibration pulses. 
To this end the modules are configured with a default configuration. A high threshold is chosen to avoid 
a high noise occupancy. A number of calibration pulses inject charge into some tracker strips and the tracker modules are read out. The readout data is scanned
on the fly for missing data and any error words from the TRB. Then the recorded data is analysed 
to show that hits are seen in the expected strips and that no errors from the modules are received.

\subsubsection{Performance test}

Two performance aspects are interesting for the TRB. Firstly, the speed with which calibration triggers 
can be sent to the TRB. Secondly, the maximum data throughput of event data from the TRB to the 
readout computer. 


\paragraph{Data throughput:}
The theoretically possible raw data rate is given by the transfer speed on the GPIO board. 
For UDP this is 1~Gbit/s. 
The UDP protocol via IPv4 Ethernet has 66 bytes of header information in every UDP packet. The minimum size of a TRB event is 18 bytes, which leads to a minimum data rate of 26.8~MBytes/s. The TRB  fills the UDP packet with all available data to the maximum payload size of 1472 bytes, which yields a maximum data rate of 120~MBytes/s.

The expected data rate at 0.04\% occupancy typical during data taking is 18 bytes per module per event, yielding 150 bytes 
of module data per TRB per event.
Including the TRB event information and overhead in the DAQ protocol
the expected data rate is about 0.212 kBytes/event. With the worst case scenario trigger rate of 2 kHz this amounts to a 
sustained data rate of only 424 kBytes/s which is well below the limitations of the transport layers.
Calibrations scans on the other hand lead to an occupancy of typically 25\% of a module, corresponding to 288 bytes of data per module and an event size of 8.5~kBytes/event.\footnote{Due to data compression on the SCT chip, the module data size does not scale linearly with hit multiplicity.}

The maximum achievable data throughput from the TRB to the readout PC has been measured by issuing repeated 
bursts of triggers of 1k events. Charge was injected into 1/4 of all strips to generate hits.  
The delay between the triggers was adjusted automatically by the TRB firmware to avoid an overflow of the 
data FIFO. The maximum recorded data rate was 51~MBytes/s with a trigger rate regulated to 6~kHz, limited by the rate at which data can be shipped from the TRB. During standard calibration scans, the TRB top rate is 10~kHz, achieved during a noise occupancy scan reading out 0.200 bytes/event.

\subsection{DAQ software and server performance tests}
\label{sec:DAQ_com}
To test the data flow through the DAQ applications, in particular the event builder and file writer applications, as well as the performance of the DAQ server, a set of simple front-end emulators and receivers is created. The emulators allow the generation of test data synchronously at an adjustable rate. These data are sent to the corresponding receivers using regular UDP packets. They are subsequently transmitted to the event builder and beyond for assembly and recording. The size of each test data fragment can be adjusted to match that expected in data taking or for testing different aspects of the data flow.

The emulators have been used throughout the DAQ developments for testing and performance tuning. A final set of performance measurements is run on the production DAQ server to confirm that the data taking will not be limited by the DAQ applications or the single DAQ server. We use the spare DAQ server to run fourteen front-end emulators, to emulate the full system connections (12 TRBs, 1 digitizer and 1 TLB). Each emulator has realistic fragment sizes leading to a total of about 20~kBytes/event. It is observed that it is possible to saturate the 1~Gbit/s Ethernet link between the servers with 5~kHz event rate without being limited by the DAQ processes. The busiest DAQ thread measures 65\% single-thread utilization and corresponds to the one running the core thread of the event builder. The full event builder process runs at 200\% multi-threaded single-core utilization and the file writer runs at 80\%. This demonstrates that the limiting factor will be the network link and not the DAQ server. Reducing the size of the emulator for the digitizer to avoid the network link limitation, we obtain a maximum rate of 12~kHz; in this case, the event builder is no longer able to sustain the rate as the fragment queue in the connection to the event builder saturates. With the number of front-end emulators reduced from fourteen to eight the system is able to reach 20~kHz, beyond which the fragment drop rate to the event builder becomes nonzero. This indicates that further throughput could be reached by merging fragments in multiple steps, though this will not be needed for the FASER experiment. Finally, running with two front-end emulators on the production DAQ server to avoid network and event builder limitations, the file writer limit is reached at a recording rate of 235~MBytes/s.




\subsection{Combined system measurements}\label{sec:combined_com}


The TDAQ system is thoroughly stress tested with the complete detector connected after the installation of the FASER components in TI12. 
The PMT signals at each scintillator or calorimeter station are mapped to a final trigger item output of the TLB using a combination of digitizer logic and TLB trigger LUT logic, as indicated in table~\ref{tab:pmt_trig_mapping}. A trigger fires in the case of signals in any calorimeter module, in the top or bottom part of the timing station, in either of the veto layers or in the preshower layer.  

\begin{table}[t]
\begin{center}
 \caption{Mapping of the PMT signals to final TLB trigger item. The PMT name indicates the source of the signal. The names are not introduced or explained, they can only be treated in the context of this paper as separate signals.}
 \begin{tabular}{||c|p{0.15\linewidth}|p{0.15\linewidth}|p{0.15\linewidth}|p{0.15\linewidth}|} 
 \hline
 PMT & \centering Digitizer Logic & \centering TLB Trigger Line & \centering TLB Trigger Logic & {\centering TLB Trigger Item} \\ [0.5ex] 
 \hline\hline
  Calo Bottom PMT 1 &  &  & &\\
 \cline{0-0}
  Calo Bottom PMT 2 & OR & 0 &  & \\
  \cline{0-2}
  Calo Top PMT 1 &  &  & & \\
 \cline{0-0}
  Calo Top PMT 2 & OR & 1 & OR & 0 \\
 \hline
  Veto 1st Layer PMT 1 & & & & \\
 \cline{0-0}
  Veto 1st Layer PMT 2 & AND & 2 & & \\
 \cline{0-2}
  Veto 2nd Layer PMT 1 & & & & \\
 \cline{0-0}
  Veto 2nd Layer PMT 2 & AND & 3 & OR & 1 \\
 \hline
  Timing Top Left PMT & & & & \\
 \cline{0-0}
  Timing Top Right PMT & AND & 4 & & \\
 \cline{0-2}
  Timing Bottom Left PMT & & & & \\
 \cline{0-0}
  Timing Bottom Right PMT & AND & 5 & OR & 2 \\
 \hline
  Preshower PMT 1 &  &  &  &  \\
 \cline{0-0}
  Preshower PMT 2 & AND & 6 & -- & 3 \\
 \hline
 \end{tabular}
 \label{tab:pmt_trig_mapping}
 \end{center}
 \end{table}
 
\begin{figure}[b]%
  \centering
  \includegraphics[width=0.98\textwidth]{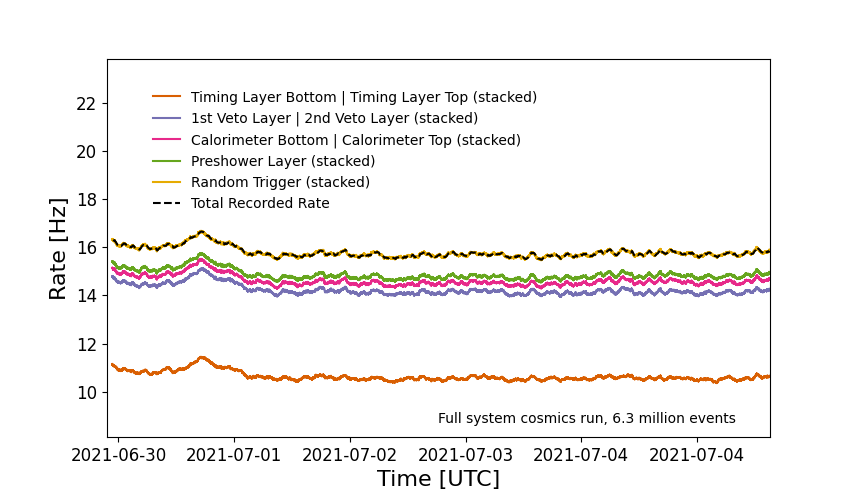}
  \caption{Individual (stacked) and total trigger rates for a multi-day full system cosmics data taking run. Each physics trigger item is mapped to a different scintillator or calorimeter station. A $\sim$1 Hz prescaled random trigger is also included.}
  \label{fig:tav_rates_cosmics_run}
\end{figure}

The rate breakdown of each trigger item is shown in figure~\ref{fig:tav_rates_cosmics_run}. The final total rate is around 16~Hz of cosmics / background noise events. All 9 tracker planes of the main part of the FASER detector (i.e. excluding the interface tracker for FASER$\nu$) are active.
The initial step to commissioning the complete integrated TDAQ system is the timing in of all sub-components to ensure that each component is reading out data from the same \BC{} (coarse timing) and that the data is well-sampled in the readout window (fine timing). Following that, the full system is stress tested by running at various high rates as well as running in cosmics mode whenever possible.

\subsubsection{Time tuning of combined system}

Both the timing for the PMT and tracker components need to be tuned for data taking for a synchronised system.

\begin{figure}[t]%
      \centering
    \includegraphics[width=0.5\textwidth]{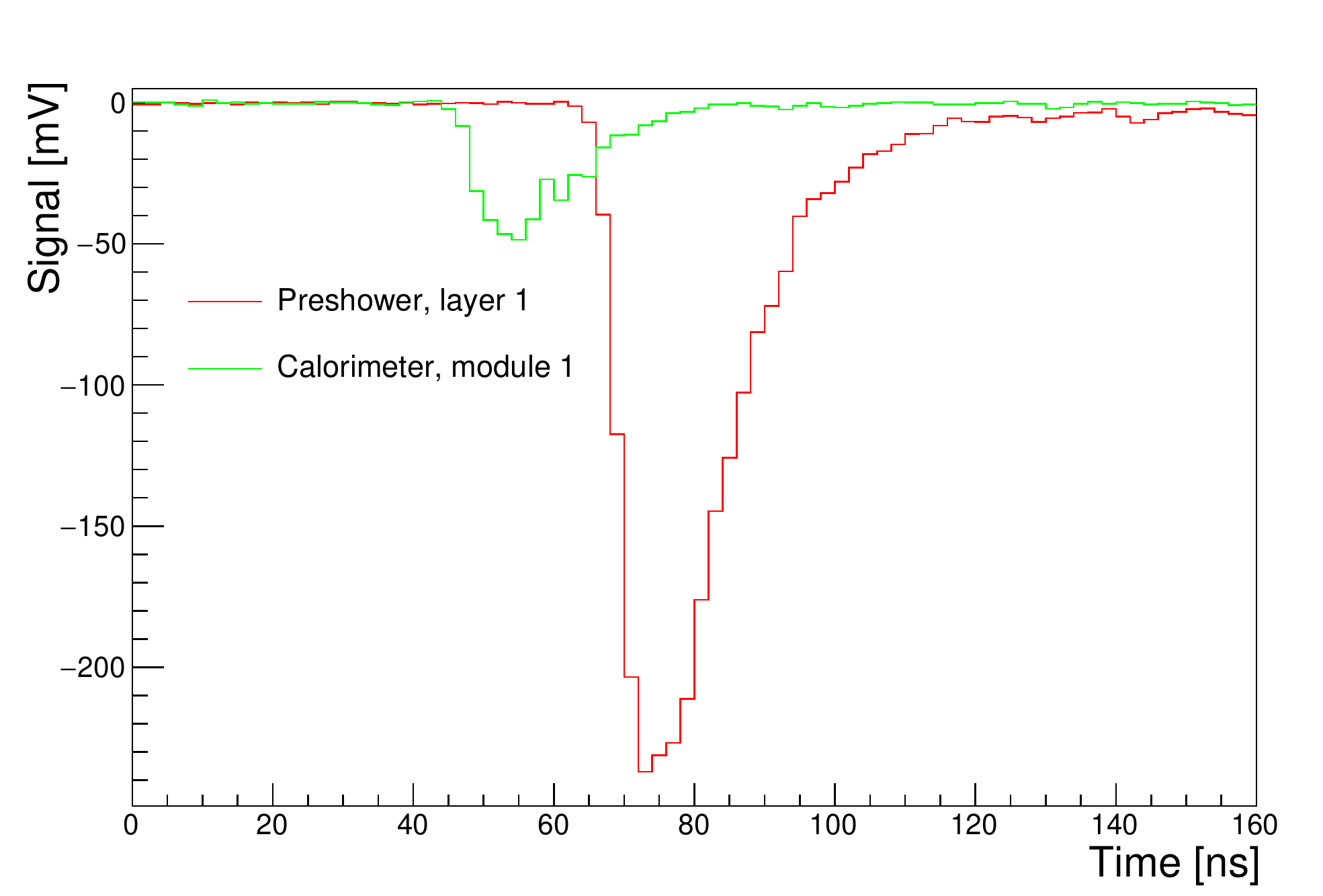}
  
  \caption{Signal waveforms as digitized by the digitizer during a cosmic ray event with a muon passing through both the preshower station (red) and a calorimeter module (green). Cable lengths are identical, but the calorimeter signal arrives around 20~ns earlier than the preshower due to the difference in PMT transition time (roughly 23~ns) as well as time-of-flight differences (roughly 2~ns). }
  \label{fig:digitizerTiming}
\end{figure}

The arrival times of PMT signals in the final system are initially not synchronised, evidenced by a comparison of the waveforms recorded by the digitizer in figure \ref{fig:digitizerTiming}. The calorimeter pulses arrive around 20~ns earlier with respect to the scintillator signals on account of differing PMT transition times and time-of-flight, and thus always trigger the event. The 1 BC misalignment can also be detected in TLB physics event data which encodes detected input signal bits in both the current and following bunch crossing. An input delay of 1 BC is set on the TLB for the calorimeter input channels to synchronise all trigger channels.

In the case of the tracker, a coarse timing step is performed to tune the arrival of the L1A/BCR signal to the SCT modules so as to adjust for the tracker readout internal pipeline of 132 \BCs{} described in section \ref{sec:hw_trb}.
The inherent signal propagation delay reduces the additional time that the L1A signal is held back before being transmitted to the SCT modules: 

\begin{align*}     
\mathrm{D}_{\mathrm{L1A}} \text{[BCs]} &= 132 - 1\; \text{(PMT transition)}\\
    & - 16\; \text{(TLB+digitizer signal processing)}\\
    & - 5\; \text{(TRB L1A to SCT modules)} \\
    & - 7\; \text{(SCT L1A decoding + SCT latency)} \\
    & - 3\; \text{(cable latency)} \\
    &= 100 \text{ BCs} 
\end{align*}


For each tracker hit the data encodes a `hit pattern': a 3-bit word corresponding to whether the hit signal was over threshold in each of the three bunch crossings that are read out. The first bit corresponds to the \BC{} before the L1A, the second bit is the same \BC{} as the L1A and the third is the \BC{} after. For the system to be correctly timed in, hits corresponding to tracks from a collision event should follow a `01X' hit pattern: nothing in the first bit, over threshold in the second bit, and no requirement on the third bit.

\begin{figure}[t]%
  \begin{subfigure}[3~ns offset]{
    \centering
    \includegraphics[width=0.5\textwidth]{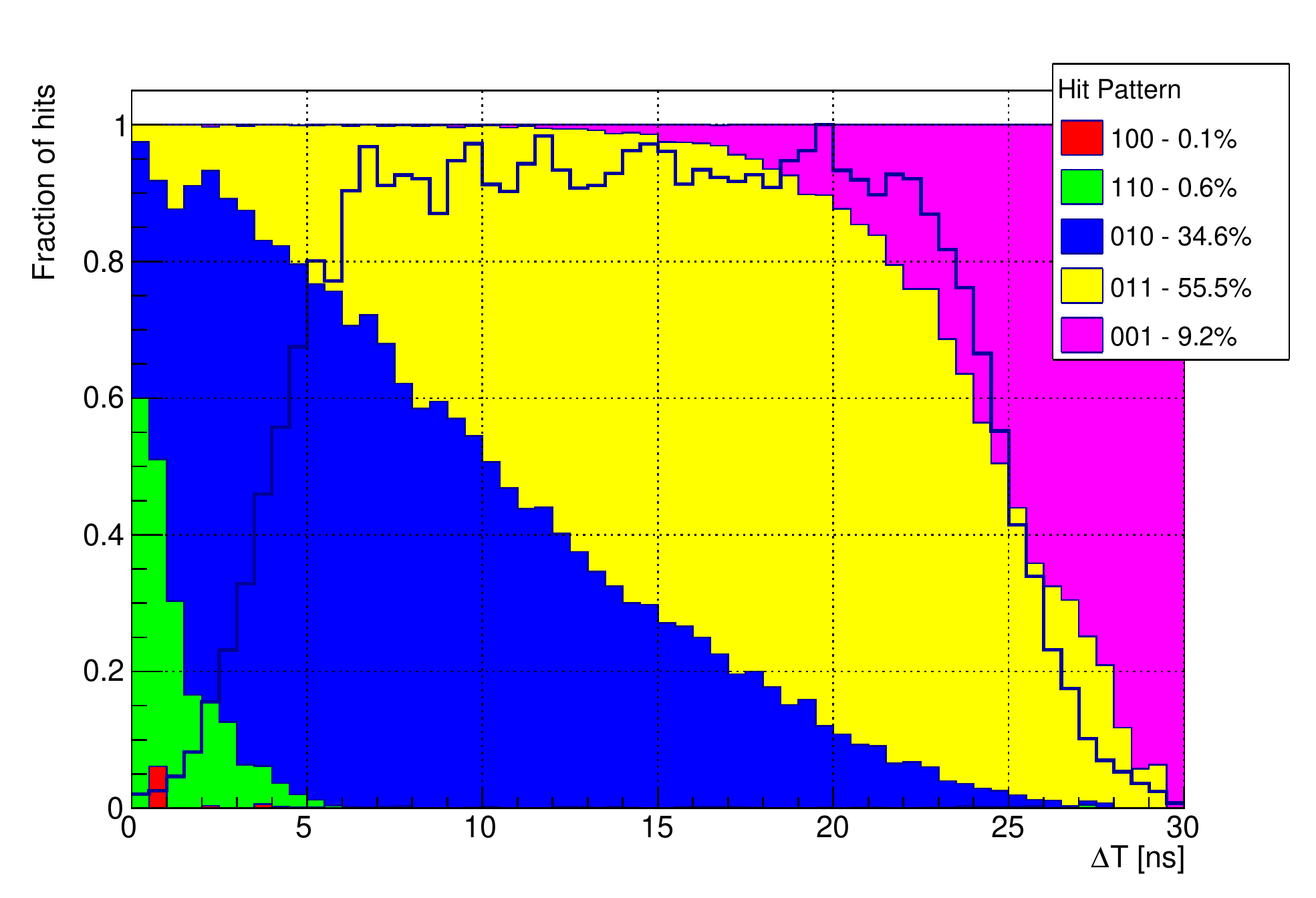}
    }
  \end{subfigure}%
    \begin{subfigure}[6~ns offset]{
    \centering
    \includegraphics[width=0.5\textwidth]{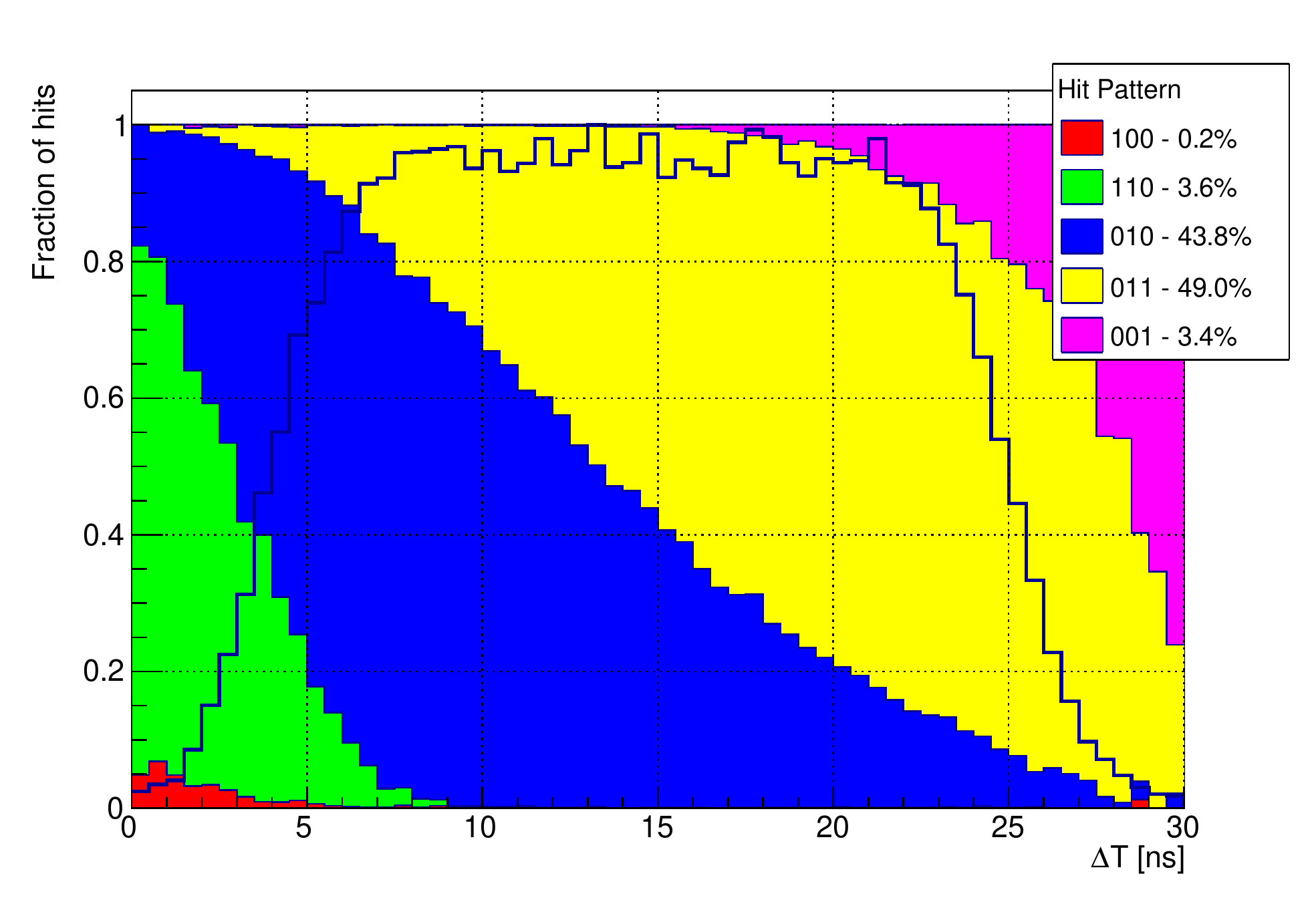}
    }
  \end{subfigure}%
    \begin{subfigure}[12.5~ns offset]{
    \centering
    \includegraphics[width=0.5\textwidth]{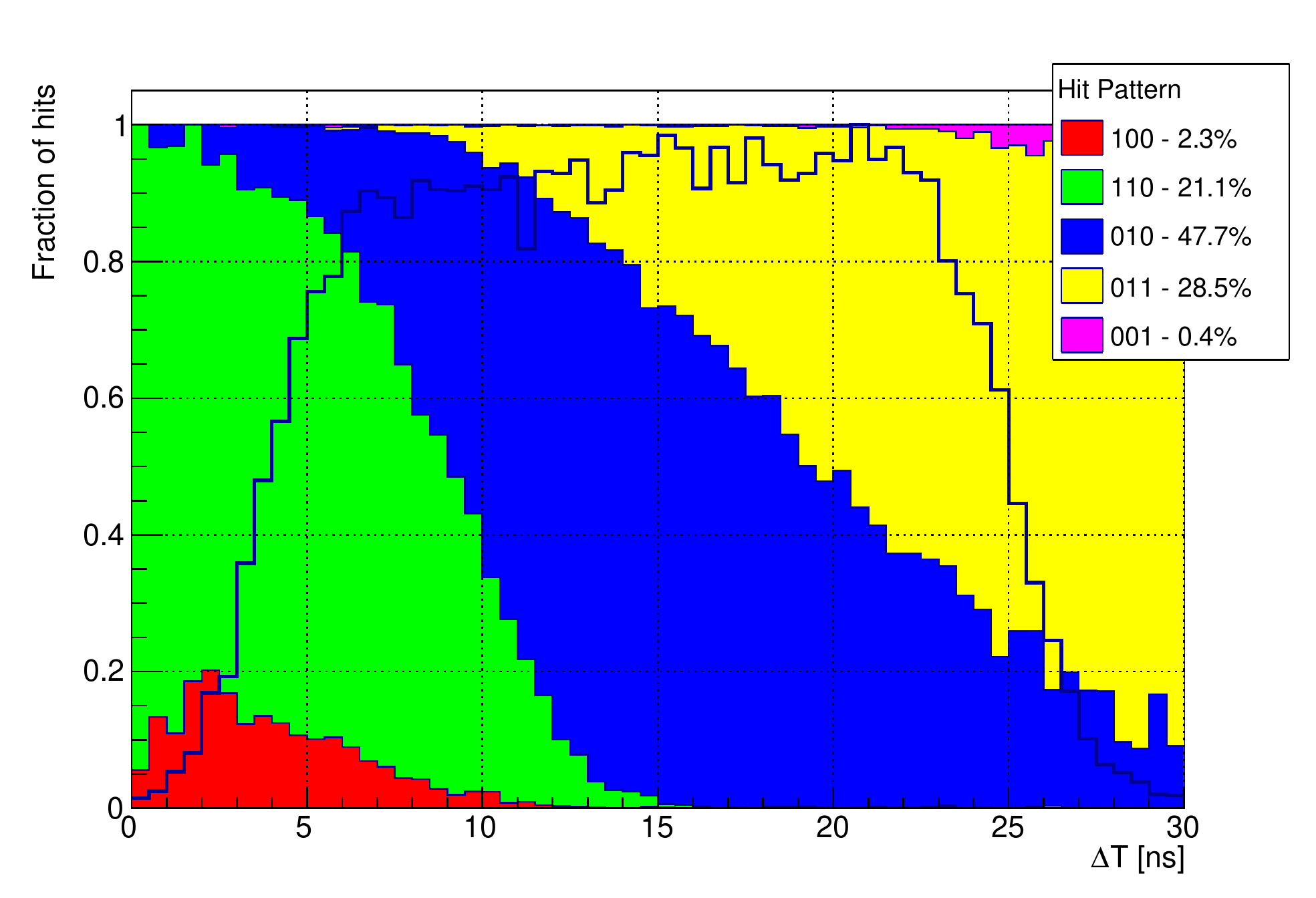}
    }
  \end{subfigure}%
    \begin{subfigure}[19~ns offset]{
    \centering
    \includegraphics[width=0.5\textwidth]{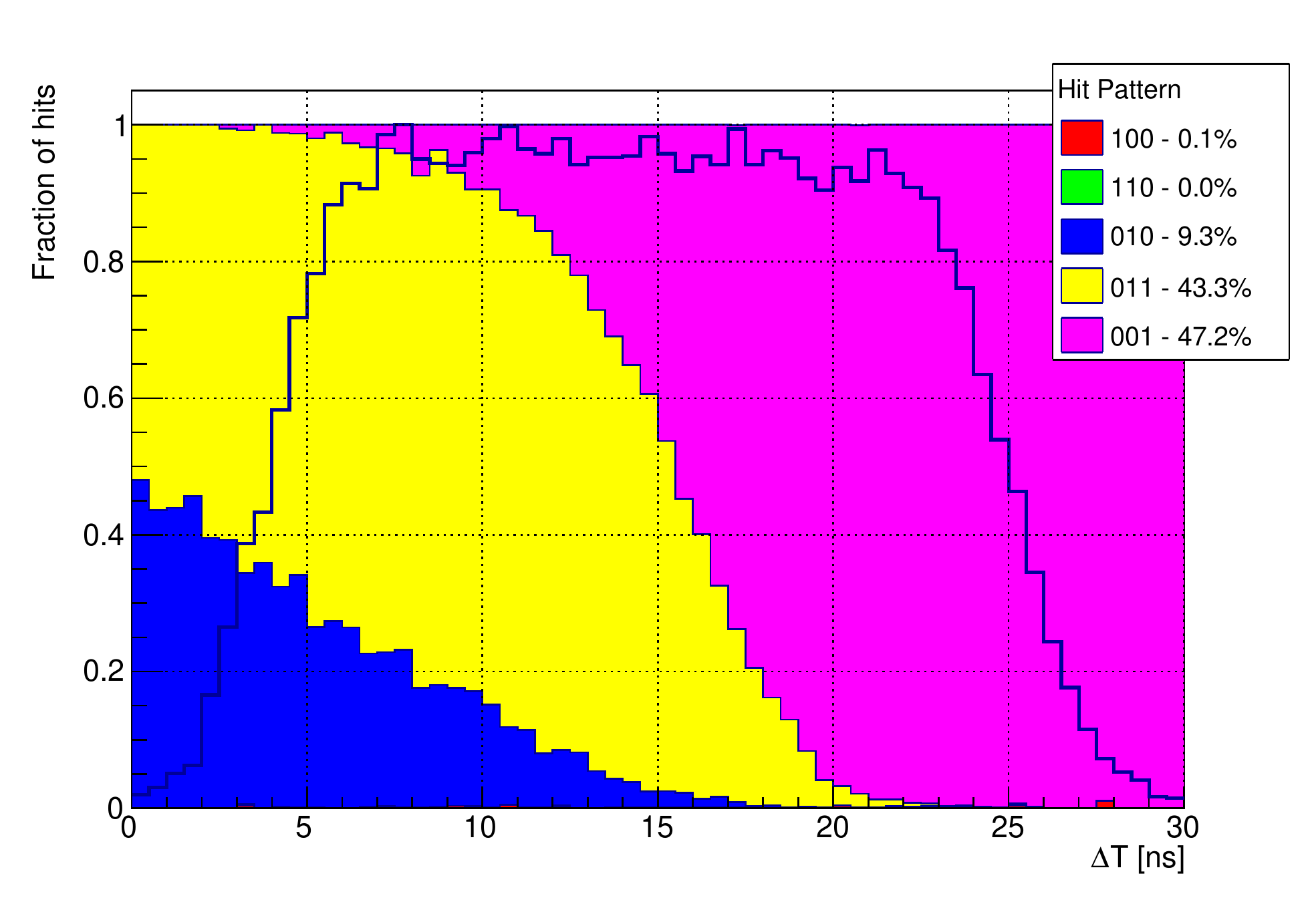}
    }
    \end{subfigure}%
  \caption{Stacked distribution of track hit patterns (3 BCs readout window) in cosmic ray data as a function of the timing of the triggering PMT pulse with respect to the system clock. The distribution of the timing is shown with the continuous dark line. The system clock is digitized at the digitizer and read out together with the PMT pulse. Four different track fine time settings are shown with a common coarse delay of 100~\BCs{}. In the final system, tracker hits from particles originating from physics collisions are expected to be narrowly centered around a fixed $\Delta$T and the tracker fine timing tuned to maximise the `01X' hit patterns for such hits.}
  \label{fig:trk_hit_pattern_fine_time}
\end{figure}

The final coarse delay for tracker data is determined by measuring the hit patterns in cosmic ray data and tuning the delay until the largest fraction of events exhibit a hit in the central \BC{} in the 3 \BCs{} hit readout window. 
As predicted, a final baseline tuning of 100 BCs proved correct for centering hits in the 3 BCs readout window. This is demonstrated in figure \ref{fig:trk_hit_pattern_fine_time}.

A tracker hit can be more precisely centered via a 390 ps fine time adjustment on the TRB described in section \ref{sec:hw_trb}.
This fine timing adjustment functionality can already be verified using the timing of cosmic ray signals with respect to the (LHC) system clock cycle. This is demonstrated in figure~\ref{fig:trk_hit_pattern_fine_time}, where the majority hit pattern is shown to shift with fine time adjustment at constant system clock phase. This figure also shows that we can achieve a fine timing that guarantees `01X' hit patterns for a physics signal synchronised to the LHC clock.

The final fine-time tuning for data taking will be performed with high energy muons from the proton-proton collisions synchronised to the LHC clock. The detector will be tuned relative to the timing scintillator station. The incoming particle time-of-flight will be adjusted for at each tracker layer by tuning the L1A coarse time delay as well as fine time delay of each TRB. This will be determined via an analysis of first collision data: Following a scan over a viable fine time range, the fine time that maximises centred `01X' hit patterns for hits on good quality reconstructed tracks will be adopted.

Once the event data for each subdetector is verified to be in the center of the readout window, it is further cross-checked that all fragments from a common event are assigned the same BCID. Since the digitizer runs at its own clock frequency, the calculation of its BCID is only accurate to within $\pm$ 1 BC, while the TRB and TLB BCIDs are expected to match exactly. Due to different internal processing delays on the TRB, its BCID in firmware is offset by a constant 9 BCs with respect to the TLB. This is corrected for in the TRB receiver software. During data taking, fragments are combined through a common event ID but the BCIDs are verified to agree within tolerance, and  result in a real-time alert and an event error flag if they differ. 



\subsubsection{TDAQ system stress testing}

A FASER physics trigger rate of 500--1000~Hz is expected during proton-proton data taking. It is therefore imperative that the full integrated system is stress tested at much higher rates than observed during cosmics runs. This section presents the current performance and bottlenecks of hardware and software readout of the complete integrated FASER TDAQ system. 

\begin{figure}[t]%
  \centering
  \includegraphics[width=0.98\textwidth]{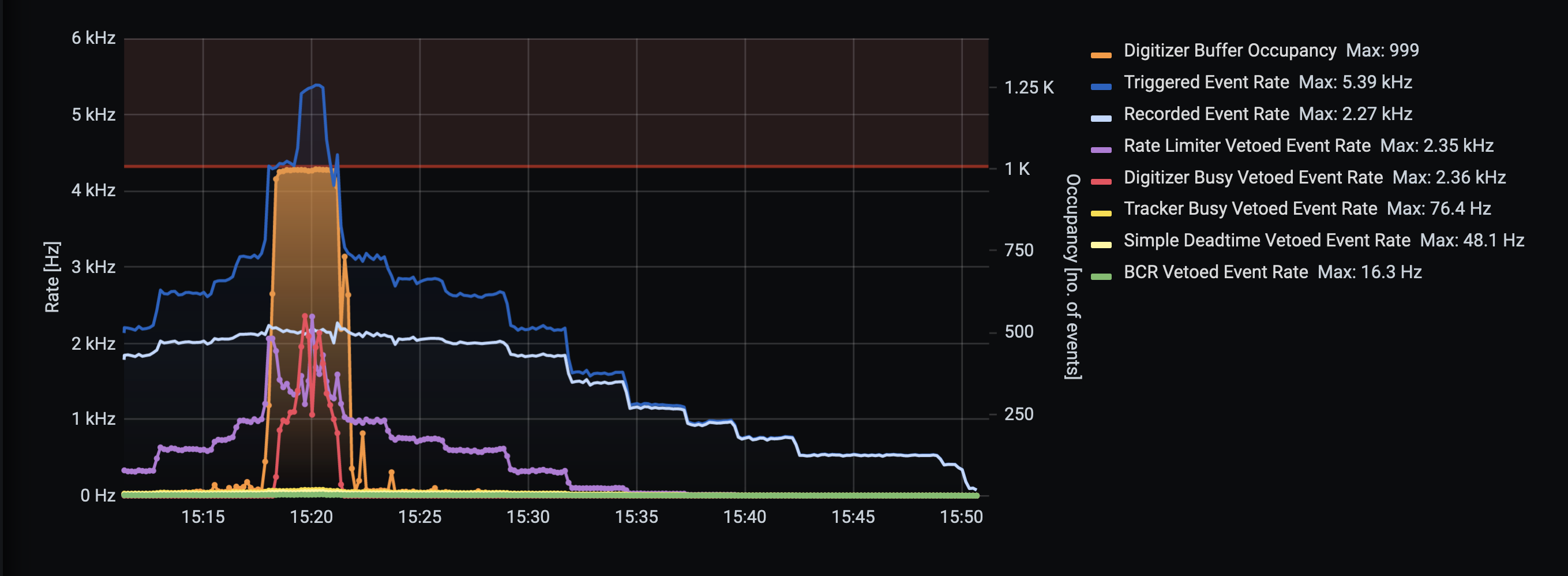}
  \caption{A stress test run with increasing input trigger rate to probe the system readout limitations and hardware component functionality in regulating rate. The triggered and recorded event rates are shown, along with the vetoed event rate due to the digitizer busy assertion, tracker busy assertion and rate limiter. The digitizer readout buffer event occupancy is shown for comparison (right y-axis). The digitizer buffer threshold is 1000 events.}
  \label{fig:graf_highrate_rate_vs_digi_occupancy}
\end{figure}

Figure~\ref{fig:graf_highrate_rate_vs_digi_occupancy} displays trigger and veto rates during a high rate test with all connected digitizer readout channels enabled (maximum data throughput), demonstrating the system functionality of regulating rate to avoid hardware components toppling over. To achieve the range in input trigger rates, a noise signal from a 
pulse generator was used and its amplitude adjusted to generate random signals at rates between 250~Hz to 5.3~kHz.

The hardware trigger system automatically regulates the rate via the TLB rate limiter and busy signals asserted by the digitizer in the case of an overfilled readout buffer, and by the TRBs  during event readout. The digitizer was set to its minimum performance configuration of one event per readout.
The maximum recorded event rate achieved was 2.2~kHz - limited by the rate limiter as well as digitizer. 
At rates up to 900 Hz, the deadtime is dominated by the tracker busy signal contributing up to a fraction of 0.7\%. The rate limiter is the dominant source of deadtime from 900~Hz to 4~kHz up to a fraction of 42\%. Once the run hits the maximum rate limiter allowed output rate, the digitizer begins to build up to a constant deadtime, though this can be removed by reading out two events per request. Final hardware performance metrics during the stress test run of figure~\ref{fig:graf_highrate_rate_vs_digi_occupancy} are summarised in table~\ref{tab:DAQ_phys_metrics}.

\begin{table}[h]
\centering
\caption{Hardware Readout performance metric measurements from the stress test run at various input rates and simple deadtime window of 150 BCs is shown in figure~\ref{fig:graf_highrate_rate_vs_digi_occupancy}. Lost events and BCID mismatches are measured to be zero in all cases. Those as well as the global deadtime are the metrics that are aimed to be minimised in these tests.\\}%
\begin{tabular}{|p{0.23\linewidth}|p{0.19\linewidth}|p{0.24\linewidth}|p{0.19\linewidth}|}
\hline
\textbf{Avg input trigger rate} & \textbf{Global deadtime} & \textbf{Max deadtime source} & \textbf{Data throughput} \\ [0.5ex]
\hline
500 Hz  & 0.9\% & tracker & 10 MBytes/s \\
\hline
1.2 kHz  & 3.5\% & rate limiter & 26 MBytes/s  \\
\hline
2.2 kHz  & 16.4\% & rate limiter & 47 MBytes/s  \\
\hline
3.2 kHz  & 33.8\% & rate limiter & 47 MBytes/s  \\
\hline
5.3 kHz  & 59.4\% & digitizer & 47 MBytes/s  \\
\hline
\end{tabular}
\label{tab:DAQ_phys_metrics}
\end{table}

The FASER DAQ software has proven itself to handle the full integrated TDAQ system up to the maximum hardware readout limited rate of 2.2~kHz and throughput of 47~MBytes/s. 
The FASER DAQ software runs 26 separate modules in parallel: the digitizer and TLB receivers, 9 tracker receivers, 1 event builder, 1 file writer and finally 13 monitoring applications. The configuration time at run start is 25 seconds, limited by the tracker receivers which set the threshold configurations for each strip of each chip of the connected tracker plane in sequence.
The maximum CPU consumer is the event builder while sending data out via the different output streams to both the file writer and monitoring applications.
The digitizer receiver application is the next highest CPU consumer, but at less than half the event builder CPU. The memory usage for each application during high rate tests and after multiple days of running in cosmics mode, is found to remain level at around the 0.1\% level. 

\begin{figure}[t]%
  \centering
  \includegraphics[width=0.98\textwidth]{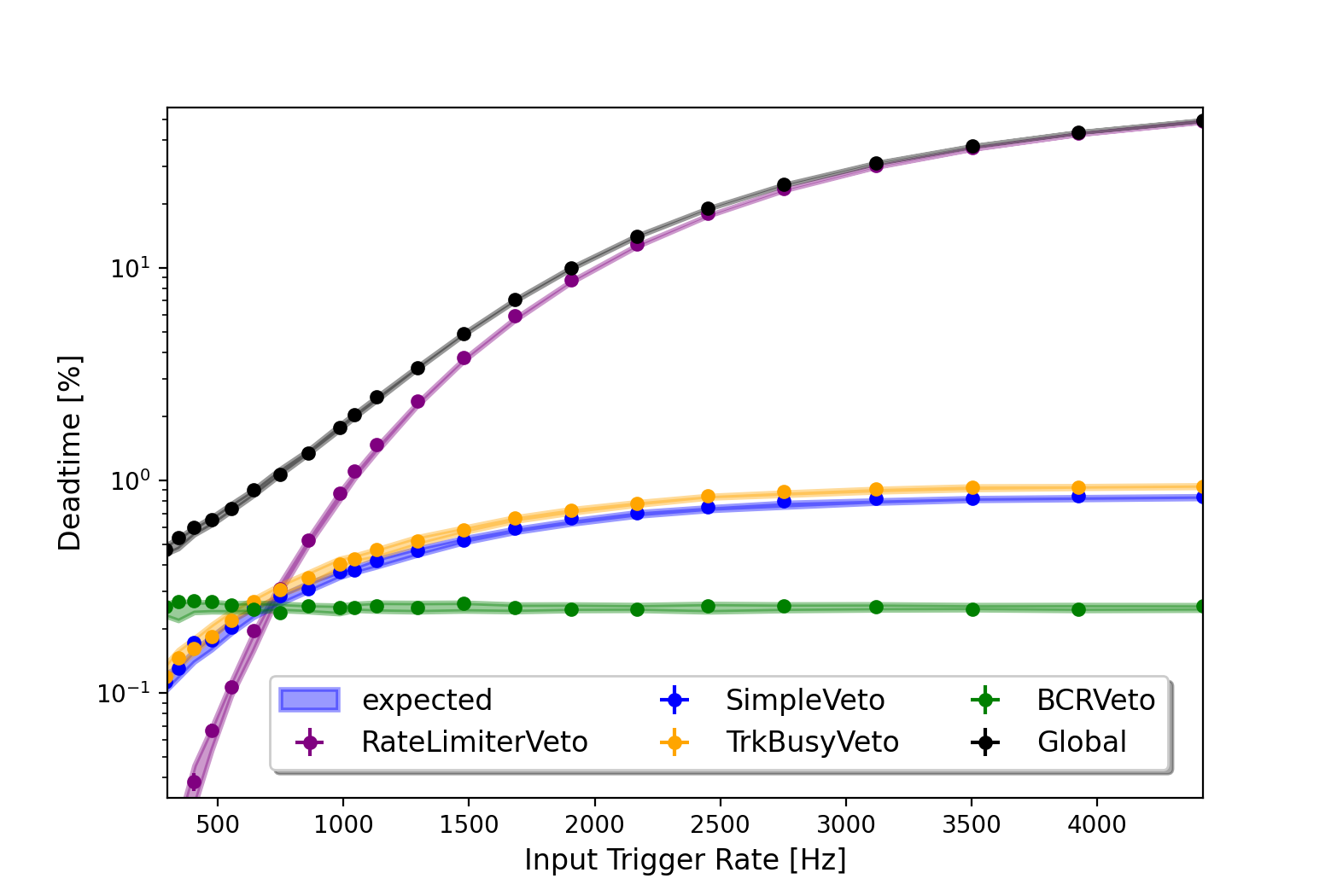}
  \caption{The measured deadtime fraction in percentage due to various sources as a function of input trigger rate is shown for a combined system run in a lab setup. Random triggers were generated using the noise signal of a signal generator as described in the text. Bands indicate the expected deadtime generated by running a trigger simulation. 
  The simple deadtime window setting was 150 BCs, and the tracker busy duration was measured to be 168 BCs.}
  \label{fig:ehn1_rate_vs_deadtime}
\end{figure}

The system deadtime as a function of rate was verified in a lab setup using an advanced pulse generator to achieve a Poissonian distribution of input trigger signals. 
The lab setup used 
three tracker readout boards connected to a tracker plane each, as well as the spare TLB and digitizer boards.
The measured deadtime of each source is shown in figure~\ref{fig:ehn1_rate_vs_deadtime}. The coloured bands show the fluctuating range of predicted deadtime, 
for a simple deadtime window set to 150 BCs.
The measured data points are found to be in agreement with prediction.
For our triggering range of interest of 500 -- 1000~Hz, the global deadtime varies from 0.7\% to 1.8\%. The rate limiter becomes the dominant source of deadtime at a rate of around 775 Hz and a rate limiter deadtime of 0.35\%. 

\newpage
\section{Summary}

The FASER experiment at the CERN LHC is getting ready for data taking in Run 3. For smooth operations and highly efficient data collection, a robust TDAQ system is required. The system has been developed and tested in standalone and combined modes. 
The hardware and software of the FASER TDAQ system is proven to run robustly: the system can sustain stable running for multiple days and at high rates without the loss of events or mis-assigned event fragments.
The system achieves a maximum performance of an output trigger rate of 2.2~kHz, corresponding to a data throughput of 47~MBytes/s. The primary limit is a built-in rate-limiter. In the current setup the digitizer would be the next limiting factor, unless the readout digitizer window, and therefore the event size, is reduced. 
The DAQ software operates stably with 26 processes running in parallel. The highest CPU consumer is the event builder while sending data over all output streams. 
The commissioning work will continue with further stress-testing and long-term runs with cosmics data taking, until the physics data taking starts in 2022.



\section*{Acknowledgments}
We thank the technical and administrative staff members at all FASER institutions for their contributions to the success of the FASER effort. The FASER collaboration is grateful to the CERN EP-DT team for their contributions and assistance to the DAQ software developments, to the CERN SY-BI group for their support in setting up the BOBR card and to the CERN IT-CS group for their assistance with the networking. 
We thank the ATLAS SCT Collaboration for donating spare SCT modules to FASER, the LHCb Collaboration for the loan of the calorimeter modules, and the ATLAS TDAQ Collaboration for the loan of old HLT servers which have been used by FASER. 
This work is supported in part by Heising-Simons Foundation Grant Nos. 2018-1135, 2019-1179,  2020-1840, Simons Foundation Grant No. 623683, U.S. National Science Foundation Grant Nos. PHY-2111427, PHY-2110929, and PHY-2110648, JSPS KAKENHI Grant Nos. JP20H01919, JP20K04004, and JP21H00082, and by the Swiss National Science Foundation.

\bibliography{references}

\end{document}